 \documentclass{ecai} 

\usepackage{latexsym}
\usepackage{amssymb}
\usepackage{amsmath}
\usepackage{amsthm}
\usepackage{booktabs}
\usepackage{graphicx}
\usepackage{color}
\usepackage[inline]{enumitem}

\usepackage{marginnote}
\usepackage{multirow}
\usepackage{paralist}
\usepackage{pifont}
\usepackage{placeins}
\usepackage{rotate} 
\usepackage[disable]{todonotes}
\usepackage{tikz}
\usepackage{times} 
\usepackage{url}
\usepackage{varwidth}
\usepackage{verbatim} 
\usepackage{wrapfig}
\usepackage{xcolor,colortbl}
\usepackage{xspace}
\usepackage{subcaption}
\usetikzlibrary{calc}
\usepackage{cleveref}

\newtheorem{theorem}{Theorem}

\newtheorem{proposition}[theorem]{Proposition}

\newtheorem{definition}{Definition}
\newtheorem{remark}{Remark}
\newtheorem{example}[theorem]{Example}

\definecolor {snow}                {rgb}{1.00,0.98,0.98}
\definecolor {ghostwhite}          {rgb}{0.97,0.97,1.00}
\definecolor {whitesmoke}          {rgb}{0.96,0.96,0.96}
\definecolor {gainsboro}           {rgb}{0.86,0.86,0.86}
\definecolor {floralwhite}         {rgb}{1.00,0.98,0.94}
\definecolor {oldlace}             {rgb}{0.99,0.96,0.90}
\definecolor {linen}               {rgb}{0.98,0.94,0.90}
\definecolor {antiquewhite}        {rgb}{0.98,0.92,0.84}
\definecolor {papayawhip}          {rgb}{1.00,0.94,0.84}
\definecolor {blanchedalmond}      {rgb}{1.00,0.92,0.80}
\definecolor {bisque}              {rgb}{1.00,0.89,0.77}
\definecolor {peachpuff}           {rgb}{1.00,0.85,0.73}
\definecolor {navajowhite}         {rgb}{1.00,0.87,0.68}
\definecolor {moccasin}            {rgb}{1.00,0.89,0.71}
\definecolor {cornsilk}            {rgb}{1.00,0.97,0.86}
\definecolor {ivory}               {rgb}{1.00,1.00,0.94}
\definecolor {lemonchiffon}        {rgb}{1.00,0.98,0.80}
\definecolor {seashell}            {rgb}{1.00,0.96,0.93}
\definecolor {honeydew}            {rgb}{0.94,1.00,0.94}
\definecolor {mintcream}           {rgb}{0.96,1.00,0.98}
\definecolor {azure}               {rgb}{0.94,1.00,1.00}
\definecolor {aliceblue}           {rgb}{0.94,0.97,1.00}
\definecolor {lavender}            {rgb}{0.90,0.90,0.98}
\definecolor {lavenderblush}       {rgb}{1.00,0.94,0.96}
\definecolor {mistyrose}           {rgb}{1.00,0.89,0.88}
\definecolor {white}               {rgb}{1.00,1.00,1.00}
\definecolor {black}               {rgb}{0.00,0.00,0.00}
\definecolor {darkslategray}       {rgb}{0.18,0.31,0.31}
\definecolor {dimgray}             {rgb}{0.41,0.41,0.41}
\definecolor {slategray}           {rgb}{0.44,0.50,0.56}
\definecolor {lightslategray}      {rgb}{0.47,0.53,0.60}
\definecolor {gray}                {rgb}{0.75,0.75,0.75}
\definecolor {lightgrey}           {rgb}{0.83,0.83,0.83}
\definecolor {midnightblue}        {rgb}{0.10,0.10,0.44}
\definecolor {navy}                {rgb}{0.00,0.00,0.50}
\definecolor {cornflowerblue}      {rgb}{0.39,0.58,0.93}
\definecolor {darkslateblue}       {rgb}{0.28,0.24,0.55}
\definecolor {slateblue}           {rgb}{0.42,0.35,0.80}
\definecolor {mediumslateblue}     {rgb}{0.48,0.41,0.93}
\definecolor {lightslateblue}      {rgb}{0.52,0.44,1.00}
\definecolor {mediumblue}          {rgb}{0.00,0.00,0.80}
\definecolor {royalblue}           {rgb}{0.25,0.41,0.88}
\definecolor {blue}                {rgb}{0.00,0.00,1.00}
\definecolor {dodgerblue}          {rgb}{0.12,0.56,1.00}
\definecolor {deepskyblue}         {rgb}{0.00,0.75,1.00}
\definecolor {skyblue}             {rgb}{0.53,0.81,0.92}
\definecolor {lightskyblue}        {rgb}{0.53,0.81,0.98}
\definecolor {steelblue}           {rgb}{0.27,0.51,0.71}
\definecolor {lightsteelblue}      {rgb}{0.69,0.77,0.87}
\definecolor {lightblue}           {rgb}{0.68,0.85,0.90}
\definecolor {powderblue}          {rgb}{0.69,0.88,0.90}
\definecolor {paleturquoise}       {rgb}{0.69,0.93,0.93}
\definecolor {darkturquoise}       {rgb}{0.00,0.81,0.82}
\definecolor {mediumturquoise}     {rgb}{0.28,0.82,0.80}
\definecolor {turquoise}           {rgb}{0.25,0.88,0.82}
\definecolor {cyan}                {rgb}{0.00,1.00,1.00}
\definecolor {lightcyan}           {rgb}{0.88,1.00,1.00}
\definecolor {cadetblue}           {rgb}{0.37,0.62,0.63}
\definecolor {mediumaquamarine}    {rgb}{0.40,0.80,0.67}
\definecolor {aquamarine}          {rgb}{0.50,1.00,0.83}
\definecolor {darkgreen}           {rgb}{0.00,0.39,0.00}
\definecolor {darkolivegreen}      {rgb}{0.33,0.42,0.18}
\definecolor {darkseagreen}        {rgb}{0.56,0.74,0.56}
\definecolor {seagreen}            {rgb}{0.18,0.55,0.34}
\definecolor {mediumseagreen}      {rgb}{0.24,0.70,0.44}
\definecolor {lightseagreen}       {rgb}{0.13,0.70,0.67}
\definecolor {palegreen}           {rgb}{0.60,0.98,0.60}
\definecolor {springgreen}         {rgb}{0.00,1.00,0.50}
\definecolor {lawngreen}           {rgb}{0.49,0.99,0.00}
\definecolor {green}               {rgb}{0.00,1.00,0.00}
\definecolor {chartreuse}          {rgb}{0.50,1.00,0.00}
\definecolor {mediumspringgreen}   {rgb}{0.00,0.98,0.60}
\definecolor {greenyellow}         {rgb}{0.68,1.00,0.18}
\definecolor {limegreen}           {rgb}{0.20,0.80,0.20}
\definecolor {yellowgreen}         {rgb}{0.60,0.80,0.20}
\definecolor {forestgreen}         {rgb}{0.13,0.55,0.13}
\definecolor {olivedrab}           {rgb}{0.42,0.56,0.14}
\definecolor {darkkhaki}           {rgb}{0.74,0.72,0.42}
\definecolor {khaki}               {rgb}{0.94,0.90,0.55}
\definecolor {palegoldenrod}       {rgb}{0.93,0.91,0.67}
\definecolor {lightgoldenrodyellow} {rgb}{0.98,0.98,0.82}
\definecolor {lightyellow}         {rgb}{1.00,1.00,0.88}
\definecolor {yellow}              {rgb}{1.00,1.00,0.00}
\definecolor {gold}                {rgb}{1.00,0.84,0.00}
\definecolor {lightgoldenrod}      {rgb}{0.93,0.87,0.51}
\definecolor {goldenrod}           {rgb}{0.85,0.65,0.13}
\definecolor {darkgoldenrod}       {rgb}{0.72,0.53,0.04}
\definecolor {rosybrown}           {rgb}{0.74,0.56,0.56}
\definecolor {indianred}           {rgb}{0.80,0.36,0.36}
\definecolor {saddlebrown}         {rgb}{0.55,0.27,0.07}
\definecolor {sienna}              {rgb}{0.63,0.32,0.18}
\definecolor {peru}                {rgb}{0.80,0.52,0.25}
\definecolor {burlywood}           {rgb}{0.87,0.72,0.53}
\definecolor {beige}               {rgb}{0.96,0.96,0.86}
\definecolor {wheat}               {rgb}{0.96,0.87,0.70}
\definecolor {sandybrown}          {rgb}{0.96,0.64,0.38}
\definecolor {tan}                 {rgb}{0.82,0.71,0.55}
\definecolor {chocolate}           {rgb}{0.82,0.41,0.12}
\definecolor {firebrick}           {rgb}{0.70,0.13,0.13}
\definecolor {brown}               {rgb}{0.65,0.16,0.16}
\definecolor {darksalmon}          {rgb}{0.91,0.59,0.48}
\definecolor {salmon}              {rgb}{0.98,0.50,0.45}
\definecolor {lightsalmon}         {rgb}{1.00,0.63,0.48}
\definecolor {orange}              {rgb}{1.00,0.65,0.00}
\definecolor {darkorange}          {rgb}{1.00,0.55,0.00}
\definecolor {coral}               {rgb}{1.00,0.50,0.31}
\definecolor {lightcoral}          {rgb}{0.94,0.50,0.50}
\definecolor {tomato}              {rgb}{1.00,0.39,0.28}
\definecolor {orangered}           {rgb}{1.00,0.27,0.00}
\definecolor {red}                 {rgb}{1.00,0.00,0.00}
\definecolor {hotpink}             {rgb}{1.00,0.41,0.71}
\definecolor {deeppink}            {rgb}{1.00,0.08,0.58}
\definecolor {pink}                {rgb}{1.00,0.75,0.80}
\definecolor {lightpink}           {rgb}{1.00,0.71,0.76}
\definecolor {palevioletred}       {rgb}{0.86,0.44,0.58}
\definecolor {maroon}              {rgb}{0.69,0.19,0.38}
\definecolor {mediumvioletred}     {rgb}{0.78,0.08,0.52}
\definecolor {violetred}           {rgb}{0.82,0.13,0.56}
\definecolor {magenta}             {rgb}{1.00,0.00,1.00}
\definecolor {violet}              {rgb}{0.93,0.51,0.93}
\definecolor {plum}                {rgb}{0.87,0.63,0.87}
\definecolor {orchid}              {rgb}{0.85,0.44,0.84}
\definecolor {mediumorchid}        {rgb}{0.73,0.33,0.83}
\definecolor {darkorchid}          {rgb}{0.60,0.20,0.80}
\definecolor {darkviolet}          {rgb}{0.58,0.00,0.83}
\definecolor {blueviolet}          {rgb}{0.54,0.17,0.89}
\definecolor {purple}              {rgb}{0.63,0.13,0.94}
\definecolor {mediumpurple}        {rgb}{0.58,0.44,0.86}
\definecolor {thistle}             {rgb}{0.85,0.75,0.85}
\definecolor {snow2}               {rgb}{0.93,0.91,0.91}
\definecolor {snow3}               {rgb}{0.80,0.79,0.79}
\definecolor {snow4}               {rgb}{0.55,0.54,0.54}
\definecolor {seashell2}           {rgb}{0.93,0.90,0.87}
\definecolor {seashell3}           {rgb}{0.80,0.77,0.75}
\definecolor {seashell4}           {rgb}{0.55,0.53,0.51}
\definecolor {antiquewhite1}       {rgb}{1.00,0.94,0.86}
\definecolor {antiquewhite2}       {rgb}{0.93,0.87,0.80}
\definecolor {antiquewhite3}       {rgb}{0.80,0.75,0.69}
\definecolor {antiquewhite4}       {rgb}{0.55,0.51,0.47}
\definecolor {bisque2}             {rgb}{0.93,0.84,0.72}
\definecolor {bisque3}             {rgb}{0.80,0.72,0.62}
\definecolor {bisque4}             {rgb}{0.55,0.49,0.42}
\definecolor {peachpuff2}          {rgb}{0.93,0.80,0.68}
\definecolor {peachpuff3}          {rgb}{0.80,0.69,0.58}
\definecolor {peachpuff4}          {rgb}{0.55,0.47,0.40}
\definecolor {navajowhite2}        {rgb}{0.93,0.81,0.63}
\definecolor {navajowhite3}        {rgb}{0.80,0.70,0.55}
\definecolor {navajowhite4}        {rgb}{0.55,0.47,0.37}
\definecolor {lemonchiffon2}       {rgb}{0.93,0.91,0.75}
\definecolor {lemonchiffon3}       {rgb}{0.80,0.79,0.65}
\definecolor {lemonchiffon4}       {rgb}{0.55,0.54,0.44}
\definecolor {cornsilk2}           {rgb}{0.93,0.91,0.80}
\definecolor {cornsilk3}           {rgb}{0.80,0.78,0.69}
\definecolor {cornsilk4}           {rgb}{0.55,0.53,0.47}
\definecolor {ivory2}              {rgb}{0.93,0.93,0.88}
\definecolor {ivory3}              {rgb}{0.80,0.80,0.76}
\definecolor {ivory4}              {rgb}{0.55,0.55,0.51}
\definecolor {honeydew2}           {rgb}{0.88,0.93,0.88}
\definecolor {honeydew3}           {rgb}{0.76,0.80,0.76}
\definecolor {honeydew4}           {rgb}{0.51,0.55,0.51}
\definecolor {lavenderblush2}      {rgb}{0.93,0.88,0.90}
\definecolor {lavenderblush3}      {rgb}{0.80,0.76,0.77}
\definecolor {lavenderblush4}      {rgb}{0.55,0.51,0.53}
\definecolor {mistyrose2}          {rgb}{0.93,0.84,0.82}
\definecolor {mistyrose3}          {rgb}{0.80,0.72,0.71}
\definecolor {mistyrose4}          {rgb}{0.55,0.49,0.48}
\definecolor {azure2}              {rgb}{0.88,0.93,0.93}
\definecolor {azure3}              {rgb}{0.76,0.80,0.80}
\definecolor {azure4}              {rgb}{0.51,0.55,0.55}
\definecolor {slateblue1}          {rgb}{0.51,0.44,1.00}
\definecolor {slateblue2}          {rgb}{0.48,0.40,0.93}
\definecolor {slateblue3}          {rgb}{0.41,0.35,0.80}
\definecolor {slateblue4}          {rgb}{0.28,0.24,0.55}
\definecolor {royalblue1}          {rgb}{0.28,0.46,1.00}
\definecolor {royalblue2}          {rgb}{0.26,0.43,0.93}
\definecolor {royalblue3}          {rgb}{0.23,0.37,0.80}
\definecolor {royalblue4}          {rgb}{0.15,0.25,0.55}
\definecolor {blue2}               {rgb}{0.00,0.00,0.93}
\definecolor {blue4}               {rgb}{0.00,0.00,0.55}
\definecolor {dodgerblue2}         {rgb}{0.11,0.53,0.93}
\definecolor {dodgerblue3}         {rgb}{0.09,0.45,0.80}
\definecolor {dodgerblue4}         {rgb}{0.06,0.31,0.55}
\definecolor {steelblue1}          {rgb}{0.39,0.72,1.00}
\definecolor {steelblue2}          {rgb}{0.36,0.67,0.93}
\definecolor {steelblue3}          {rgb}{0.31,0.58,0.80}
\definecolor {steelblue4}          {rgb}{0.21,0.39,0.55}
\definecolor {deepskyblue2}        {rgb}{0.00,0.70,0.93}
\definecolor {deepskyblue3}        {rgb}{0.00,0.60,0.80}
\definecolor {deepskyblue4}        {rgb}{0.00,0.41,0.55}
\definecolor {skyblue1}            {rgb}{0.53,0.81,1.00}
\definecolor {skyblue2}            {rgb}{0.49,0.75,0.93}
\definecolor {skyblue3}            {rgb}{0.42,0.65,0.80}
\definecolor {skyblue4}            {rgb}{0.29,0.44,0.55}
\definecolor {lightskyblue1}       {rgb}{0.69,0.89,1.00}
\definecolor {lightskyblue2}       {rgb}{0.64,0.83,0.93}
\definecolor {lightskyblue3}       {rgb}{0.55,0.71,0.80}
\definecolor {lightskyblue4}       {rgb}{0.38,0.48,0.55}
\definecolor {slategray1}          {rgb}{0.78,0.89,1.00}
\definecolor {slategray2}          {rgb}{0.73,0.83,0.93}
\definecolor {slategray3}          {rgb}{0.62,0.71,0.80}
\definecolor {slategray4}          {rgb}{0.42,0.48,0.55}
\definecolor {lightsteelblue1}     {rgb}{0.79,0.88,1.00}
\definecolor {lightsteelblue2}     {rgb}{0.74,0.82,0.93}
\definecolor {lightsteelblue3}     {rgb}{0.64,0.71,0.80}
\definecolor {lightsteelblue4}     {rgb}{0.43,0.48,0.55}
\definecolor {lightblue1}          {rgb}{0.75,0.94,1.00}
\definecolor {lightblue2}          {rgb}{0.70,0.87,0.93}
\definecolor {lightblue3}          {rgb}{0.60,0.75,0.80}
\definecolor {lightblue4}          {rgb}{0.41,0.51,0.55}
\definecolor {lightcyan2}          {rgb}{0.82,0.93,0.93}
\definecolor {lightcyan3}          {rgb}{0.71,0.80,0.80}
\definecolor {lightcyan4}          {rgb}{0.48,0.55,0.55}
\definecolor {paleturquoise1}      {rgb}{0.73,1.00,1.00}
\definecolor {paleturquoise2}      {rgb}{0.68,0.93,0.93}
\definecolor {paleturquoise3}      {rgb}{0.59,0.80,0.80}
\definecolor {paleturquoise4}      {rgb}{0.40,0.55,0.55}
\definecolor {cadetblue1}          {rgb}{0.60,0.96,1.00}
\definecolor {cadetblue2}          {rgb}{0.56,0.90,0.93}
\definecolor {cadetblue3}          {rgb}{0.48,0.77,0.80}
\definecolor {cadetblue4}          {rgb}{0.33,0.53,0.55}
\definecolor {turquoise1}          {rgb}{0.00,0.96,1.00}
\definecolor {turquoise2}          {rgb}{0.00,0.90,0.93}
\definecolor {turquoise3}          {rgb}{0.00,0.77,0.80}
\definecolor {turquoise4}          {rgb}{0.00,0.53,0.55}
\definecolor {cyan2}               {rgb}{0.00,0.93,0.93}
\definecolor {cyan3}               {rgb}{0.00,0.80,0.80}
\definecolor {cyan4}               {rgb}{0.00,0.55,0.55}
\definecolor {darkslategray1}      {rgb}{0.59,1.00,1.00}
\definecolor {darkslategray2}      {rgb}{0.55,0.93,0.93}
\definecolor {darkslategray3}      {rgb}{0.47,0.80,0.80}
\definecolor {darkslategray4}      {rgb}{0.32,0.55,0.55}
\definecolor {aquamarine2}         {rgb}{0.46,0.93,0.78}
\definecolor {aquamarine4}         {rgb}{0.27,0.55,0.45}
\definecolor {darkseagreen1}       {rgb}{0.76,1.00,0.76}
\definecolor {darkseagreen2}       {rgb}{0.71,0.93,0.71}
\definecolor {darkseagreen3}       {rgb}{0.61,0.80,0.61}
\definecolor {darkseagreen4}       {rgb}{0.41,0.55,0.41}
\definecolor {seagreen1}           {rgb}{0.33,1.00,0.62}
\definecolor {seagreen2}           {rgb}{0.31,0.93,0.58}
\definecolor {seagreen3}           {rgb}{0.26,0.80,0.50}
\definecolor {palegreen1}          {rgb}{0.60,1.00,0.60}
\definecolor {palegreen2}          {rgb}{0.56,0.93,0.56}
\definecolor {palegreen3}          {rgb}{0.49,0.80,0.49}
\definecolor {palegreen4}          {rgb}{0.33,0.55,0.33}
\definecolor {springgreen2}        {rgb}{0.00,0.93,0.46}
\definecolor {springgreen3}        {rgb}{0.00,0.80,0.40}
\definecolor {springgreen4}        {rgb}{0.00,0.55,0.27}
\definecolor {green2}              {rgb}{0.00,0.93,0.00}
\definecolor {green3}              {rgb}{0.00,0.80,0.00}
\definecolor {green4}              {rgb}{0.00,0.55,0.00}
\definecolor {chartreuse2}         {rgb}{0.46,0.93,0.00}
\definecolor {chartreuse3}         {rgb}{0.40,0.80,0.00}
\definecolor {chartreuse4}         {rgb}{0.27,0.55,0.00}
\definecolor {olivedrab1}          {rgb}{0.75,1.00,0.24}
\definecolor {olivedrab2}          {rgb}{0.70,0.93,0.23}
\definecolor {olivedrab4}          {rgb}{0.41,0.55,0.13}
\definecolor {darkolivegreen1}     {rgb}{0.79,1.00,0.44}
\definecolor {darkolivegreen2}     {rgb}{0.74,0.93,0.41}
\definecolor {darkolivegreen3}     {rgb}{0.64,0.80,0.35}
\definecolor {darkolivegreen4}     {rgb}{0.43,0.55,0.24}
\definecolor {khaki1}              {rgb}{1.00,0.96,0.56}
\definecolor {khaki2}              {rgb}{0.93,0.90,0.52}
\definecolor {khaki3}              {rgb}{0.80,0.78,0.45}
\definecolor {khaki4}              {rgb}{0.55,0.53,0.31}
\definecolor {lightgoldenrod1}     {rgb}{1.00,0.93,0.55}
\definecolor {lightgoldenrod2}     {rgb}{0.93,0.86,0.51}
\definecolor {lightgoldenrod3}     {rgb}{0.80,0.75,0.44}
\definecolor {lightgoldenrod4}     {rgb}{0.55,0.51,0.30}
\definecolor {lightyellow2}        {rgb}{0.93,0.93,0.82}
\definecolor {lightyellow3}        {rgb}{0.80,0.80,0.71}
\definecolor {lightyellow4}        {rgb}{0.55,0.55,0.48}
\definecolor {yellow2}             {rgb}{0.93,0.93,0.00}
\definecolor {yellow3}             {rgb}{0.80,0.80,0.00}
\definecolor {yellow4}             {rgb}{0.55,0.55,0.00}
\definecolor {gold2}               {rgb}{0.93,0.79,0.00}
\definecolor {gold3}               {rgb}{0.80,0.68,0.00}
\definecolor {gold4}               {rgb}{0.55,0.46,0.00}
\definecolor {goldenrod1}          {rgb}{1.00,0.76,0.15}
\definecolor {goldenrod2}          {rgb}{0.93,0.71,0.13}
\definecolor {goldenrod3}          {rgb}{0.80,0.61,0.11}
\definecolor {goldenrod4}          {rgb}{0.55,0.41,0.08}
\definecolor {darkgoldenrod1}      {rgb}{1.00,0.73,0.06}
\definecolor {darkgoldenrod2}      {rgb}{0.93,0.68,0.05}
\definecolor {darkgoldenrod3}      {rgb}{0.80,0.58,0.05}
\definecolor {darkgoldenrod4}      {rgb}{0.55,0.40,0.03}
\definecolor {rosybrown1}          {rgb}{1.00,0.76,0.76}
\definecolor {rosybrown2}          {rgb}{0.93,0.71,0.71}
\definecolor {rosybrown3}          {rgb}{0.80,0.61,0.61}
\definecolor {rosybrown4}          {rgb}{0.55,0.41,0.41}
\definecolor {indianred1}          {rgb}{1.00,0.42,0.42}
\definecolor {indianred2}          {rgb}{0.93,0.39,0.39}
\definecolor {indianred3}          {rgb}{0.80,0.33,0.33}
\definecolor {indianred4}          {rgb}{0.55,0.23,0.23}
\definecolor {sienna1}             {rgb}{1.00,0.51,0.28}
\definecolor {sienna2}             {rgb}{0.93,0.47,0.26}
\definecolor {sienna3}             {rgb}{0.80,0.41,0.22}
\definecolor {sienna4}             {rgb}{0.55,0.28,0.15}
\definecolor {burlywood1}          {rgb}{1.00,0.83,0.61}
\definecolor {burlywood2}          {rgb}{0.93,0.77,0.57}
\definecolor {burlywood3}          {rgb}{0.80,0.67,0.49}
\definecolor {burlywood4}          {rgb}{0.55,0.45,0.33}
\definecolor {wheat1}              {rgb}{1.00,0.91,0.73}
\definecolor {wheat2}              {rgb}{0.93,0.85,0.68}
\definecolor {wheat3}              {rgb}{0.80,0.73,0.59}
\definecolor {wheat4}              {rgb}{0.55,0.49,0.40}
\definecolor {tan1}                {rgb}{1.00,0.65,0.31}
\definecolor {tan2}                {rgb}{0.93,0.60,0.29}
\definecolor {tan4}                {rgb}{0.55,0.35,0.17}
\definecolor {chocolate1}          {rgb}{1.00,0.50,0.14}
\definecolor {chocolate2}          {rgb}{0.93,0.46,0.13}
\definecolor {chocolate3}          {rgb}{0.80,0.40,0.11}
\definecolor {firebrick1}          {rgb}{1.00,0.19,0.19}
\definecolor {firebrick2}          {rgb}{0.93,0.17,0.17}
\definecolor {firebrick3}          {rgb}{0.80,0.15,0.15}
\definecolor {firebrick4}          {rgb}{0.55,0.10,0.10}
\definecolor {brown1}              {rgb}{1.00,0.25,0.25}
\definecolor {brown2}              {rgb}{0.93,0.23,0.23}
\definecolor {brown3}              {rgb}{0.80,0.20,0.20}
\definecolor {brown4}              {rgb}{0.55,0.14,0.14}
\definecolor {salmon1}             {rgb}{1.00,0.55,0.41}
\definecolor {salmon2}             {rgb}{0.93,0.51,0.38}
\definecolor {salmon3}             {rgb}{0.80,0.44,0.33}
\definecolor {salmon4}             {rgb}{0.55,0.30,0.22}
\definecolor {lightsalmon2}        {rgb}{0.93,0.58,0.45}
\definecolor {lightsalmon3}        {rgb}{0.80,0.51,0.38}
\definecolor {lightsalmon4}        {rgb}{0.55,0.34,0.26}
\definecolor {orange2}             {rgb}{0.93,0.60,0.00}
\definecolor {orange3}             {rgb}{0.80,0.52,0.00}
\definecolor {orange4}             {rgb}{0.55,0.35,0.00}
\definecolor {darkorange1}         {rgb}{1.00,0.50,0.00}
\definecolor {darkorange2}         {rgb}{0.93,0.46,0.00}
\definecolor {darkorange3}         {rgb}{0.80,0.40,0.00}
\definecolor {darkorange4}         {rgb}{0.55,0.27,0.00}
\definecolor {coral1}              {rgb}{1.00,0.45,0.34}
\definecolor {coral2}              {rgb}{0.93,0.42,0.31}
\definecolor {coral3}              {rgb}{0.80,0.36,0.27}
\definecolor {coral4}              {rgb}{0.55,0.24,0.18}
\definecolor {tomato2}             {rgb}{0.93,0.36,0.26}
\definecolor {tomato3}             {rgb}{0.80,0.31,0.22}
\definecolor {tomato4}             {rgb}{0.55,0.21,0.15}
\definecolor {orangered2}          {rgb}{0.93,0.25,0.00}
\definecolor {orangered3}          {rgb}{0.80,0.22,0.00}
\definecolor {orangered4}          {rgb}{0.55,0.15,0.00}
\definecolor {red2}                {rgb}{0.93,0.00,0.00}
\definecolor {red3}                {rgb}{0.80,0.00,0.00}
\definecolor {red4}                {rgb}{0.55,0.00,0.00}
\definecolor {deeppink2}           {rgb}{0.93,0.07,0.54}
\definecolor {deeppink3}           {rgb}{0.80,0.06,0.46}
\definecolor {deeppink4}           {rgb}{0.55,0.04,0.31}
\definecolor {hotpink1}            {rgb}{1.00,0.43,0.71}
\definecolor {hotpink2}            {rgb}{0.93,0.42,0.65}
\definecolor {hotpink3}            {rgb}{0.80,0.38,0.56}
\definecolor {hotpink4}            {rgb}{0.55,0.23,0.38}
\definecolor {pink1}               {rgb}{1.00,0.71,0.77}
\definecolor {pink2}               {rgb}{0.93,0.66,0.72}
\definecolor {pink3}               {rgb}{0.80,0.57,0.62}
\definecolor {pink4}               {rgb}{0.55,0.39,0.42}
\definecolor {lightpink1}          {rgb}{1.00,0.68,0.73}
\definecolor {lightpink2}          {rgb}{0.93,0.64,0.68}
\definecolor {lightpink3}          {rgb}{0.80,0.55,0.58}
\definecolor {lightpink4}          {rgb}{0.55,0.37,0.40}
\definecolor {palevioletred1}      {rgb}{1.00,0.51,0.67}
\definecolor {palevioletred2}      {rgb}{0.93,0.47,0.62}
\definecolor {palevioletred3}      {rgb}{0.80,0.41,0.54}
\definecolor {palevioletred4}      {rgb}{0.55,0.28,0.36}
\definecolor {maroon1}             {rgb}{1.00,0.20,0.70}
\definecolor {maroon2}             {rgb}{0.93,0.19,0.65}
\definecolor {maroon3}             {rgb}{0.80,0.16,0.56}
\definecolor {maroon4}             {rgb}{0.55,0.11,0.38}
\definecolor {violetred1}          {rgb}{1.00,0.24,0.59}
\definecolor {violetred2}          {rgb}{0.93,0.23,0.55}
\definecolor {violetred3}          {rgb}{0.80,0.20,0.47}
\definecolor {violetred4}          {rgb}{0.55,0.13,0.32}
\definecolor {magenta2}            {rgb}{0.93,0.00,0.93}
\definecolor {magenta3}            {rgb}{0.80,0.00,0.80}
\definecolor {magenta4}            {rgb}{0.55,0.00,0.55}
\definecolor {orchid1}             {rgb}{1.00,0.51,0.98}
\definecolor {orchid2}             {rgb}{0.93,0.48,0.91}
\definecolor {orchid3}             {rgb}{0.80,0.41,0.79}
\definecolor {orchid4}             {rgb}{0.55,0.28,0.54}
\definecolor {plum1}               {rgb}{1.00,0.73,1.00}
\definecolor {plum2}               {rgb}{0.93,0.68,0.93}
\definecolor {plum3}               {rgb}{0.80,0.59,0.80}
\definecolor {plum4}               {rgb}{0.55,0.40,0.55}
\definecolor {mediumorchid1}       {rgb}{0.88,0.40,1.00}
\definecolor {mediumorchid2}       {rgb}{0.82,0.37,0.93}
\definecolor {mediumorchid3}       {rgb}{0.71,0.32,0.80}
\definecolor {mediumorchid4}       {rgb}{0.48,0.22,0.55}
\definecolor {darkorchid1}         {rgb}{0.75,0.24,1.00}
\definecolor {darkorchid2}         {rgb}{0.70,0.23,0.93}
\definecolor {darkorchid3}         {rgb}{0.60,0.20,0.80}
\definecolor {darkorchid4}         {rgb}{0.41,0.13,0.55}
\definecolor {purple1}             {rgb}{0.61,0.19,1.00}
\definecolor {purple2}             {rgb}{0.57,0.17,0.93}
\definecolor {purple3}             {rgb}{0.49,0.15,0.80}
\definecolor {purple4}             {rgb}{0.33,0.10,0.55}
\definecolor {mediumpurple1}       {rgb}{0.67,0.51,1.00}
\definecolor {mediumpurple2}       {rgb}{0.62,0.47,0.93}
\definecolor {mediumpurple3}       {rgb}{0.54,0.41,0.80}
\definecolor {mediumpurple4}       {rgb}{0.36,0.28,0.55}
\definecolor {thistle1}            {rgb}{1.00,0.88,1.00}
\definecolor {thistle2}            {rgb}{0.93,0.82,0.93}
\definecolor {thistle3}            {rgb}{0.80,0.71,0.80}
\definecolor {thistle4}            {rgb}{0.55,0.48,0.55}
\definecolor {gray1}               {rgb}{0.01,0.01,0.01}
\definecolor {gray2}               {rgb}{0.02,0.02,0.02}
\definecolor {gray3}               {rgb}{0.03,0.03,0.03}
\definecolor {gray4}               {rgb}{0.04,0.04,0.04}
\definecolor {gray5}               {rgb}{0.05,0.05,0.05}
\definecolor {gray6}               {rgb}{0.06,0.06,0.06}
\definecolor {gray7}               {rgb}{0.07,0.07,0.07}
\definecolor {gray8}               {rgb}{0.08,0.08,0.08}
\definecolor {gray9}               {rgb}{0.09,0.09,0.09}
\definecolor {gray10}              {rgb}{0.10,0.10,0.10}
\definecolor {gray11}              {rgb}{0.11,0.11,0.11}
\definecolor {gray12}              {rgb}{0.12,0.12,0.12}
\definecolor {gray13}              {rgb}{0.13,0.13,0.13}
\definecolor {gray14}              {rgb}{0.14,0.14,0.14}
\definecolor {gray15}              {rgb}{0.15,0.15,0.15}
\definecolor {gray16}              {rgb}{0.16,0.16,0.16}
\definecolor {gray17}              {rgb}{0.17,0.17,0.17}
\definecolor {gray18}              {rgb}{0.18,0.18,0.18}
\definecolor {gray19}              {rgb}{0.19,0.19,0.19}
\definecolor {gray20}              {rgb}{0.20,0.20,0.20}
\definecolor {gray21}              {rgb}{0.21,0.21,0.21}
\definecolor {gray22}              {rgb}{0.22,0.22,0.22}
\definecolor {gray23}              {rgb}{0.23,0.23,0.23}
\definecolor {gray24}              {rgb}{0.24,0.24,0.24}
\definecolor {gray25}              {rgb}{0.25,0.25,0.25}
\definecolor {gray26}              {rgb}{0.26,0.26,0.26}
\definecolor {gray27}              {rgb}{0.27,0.27,0.27}
\definecolor {gray28}              {rgb}{0.28,0.28,0.28}
\definecolor {gray29}              {rgb}{0.29,0.29,0.29}
\definecolor {gray30}              {rgb}{0.30,0.30,0.30}
\definecolor {gray31}              {rgb}{0.31,0.31,0.31}
\definecolor {gray32}              {rgb}{0.32,0.32,0.32}
\definecolor {gray33}              {rgb}{0.33,0.33,0.33}
\definecolor {gray34}              {rgb}{0.34,0.34,0.34}
\definecolor {gray35}              {rgb}{0.35,0.35,0.35}
\definecolor {gray36}              {rgb}{0.36,0.36,0.36}
\definecolor {gray37}              {rgb}{0.37,0.37,0.37}
\definecolor {gray38}              {rgb}{0.38,0.38,0.38}
\definecolor {gray39}              {rgb}{0.39,0.39,0.39}
\definecolor {gray40}              {rgb}{0.40,0.40,0.40}
\definecolor {gray42}              {rgb}{0.42,0.42,0.42}
\definecolor {gray43}              {rgb}{0.43,0.43,0.43}
\definecolor {gray44}              {rgb}{0.44,0.44,0.44}
\definecolor {gray45}              {rgb}{0.45,0.45,0.45}
\definecolor {gray46}              {rgb}{0.46,0.46,0.46}
\definecolor {gray47}              {rgb}{0.47,0.47,0.47}
\definecolor {gray48}              {rgb}{0.48,0.48,0.48}
\definecolor {gray49}              {rgb}{0.49,0.49,0.49}
\definecolor {gray50}              {rgb}{0.50,0.50,0.50}
\definecolor {gray51}              {rgb}{0.51,0.51,0.51}
\definecolor {gray52}              {rgb}{0.52,0.52,0.52}
\definecolor {gray53}              {rgb}{0.53,0.53,0.53}
\definecolor {gray54}              {rgb}{0.54,0.54,0.54}
\definecolor {gray55}              {rgb}{0.55,0.55,0.55}
\definecolor {gray56}              {rgb}{0.56,0.56,0.56}
\definecolor {gray57}              {rgb}{0.57,0.57,0.57}
\definecolor {gray58}              {rgb}{0.58,0.58,0.58}
\definecolor {gray59}              {rgb}{0.59,0.59,0.59}
\definecolor {gray60}              {rgb}{0.60,0.60,0.60}
\definecolor {gray61}              {rgb}{0.61,0.61,0.61}
\definecolor {gray62}              {rgb}{0.62,0.62,0.62}
\definecolor {gray63}              {rgb}{0.63,0.63,0.63}
\definecolor {gray64}              {rgb}{0.64,0.64,0.64}
\definecolor {gray65}              {rgb}{0.65,0.65,0.65}
\definecolor {gray66}              {rgb}{0.66,0.66,0.66}
\definecolor {gray67}              {rgb}{0.67,0.67,0.67}
\definecolor {gray68}              {rgb}{0.68,0.68,0.68}
\definecolor {gray69}              {rgb}{0.69,0.69,0.69}
\definecolor {gray70}              {rgb}{0.70,0.70,0.70}
\definecolor {gray71}              {rgb}{0.71,0.71,0.71}
\definecolor {gray72}              {rgb}{0.72,0.72,0.72}
\definecolor {gray73}              {rgb}{0.73,0.73,0.73}
\definecolor {gray74}              {rgb}{0.74,0.74,0.74}
\definecolor {gray75}              {rgb}{0.75,0.75,0.75}
\definecolor {gray76}              {rgb}{0.76,0.76,0.76}
\definecolor {gray77}              {rgb}{0.77,0.77,0.77}
\definecolor {gray78}              {rgb}{0.78,0.78,0.78}
\definecolor {gray79}              {rgb}{0.79,0.79,0.79}
\definecolor {gray80}              {rgb}{0.80,0.80,0.80}
\definecolor {gray81}              {rgb}{0.81,0.81,0.81}
\definecolor {gray82}              {rgb}{0.82,0.82,0.82}
\definecolor {gray83}              {rgb}{0.83,0.83,0.83}
\definecolor {gray84}              {rgb}{0.84,0.84,0.84}
\definecolor {gray85}              {rgb}{0.85,0.85,0.85}
\definecolor {gray86}              {rgb}{0.86,0.86,0.86}
\definecolor {gray87}              {rgb}{0.87,0.87,0.87}
\definecolor {gray88}              {rgb}{0.88,0.88,0.88}
\definecolor {gray89}              {rgb}{0.89,0.89,0.89}
\definecolor {gray90}              {rgb}{0.90,0.90,0.90}
\definecolor {gray91}              {rgb}{0.91,0.91,0.91}
\definecolor {gray92}              {rgb}{0.92,0.92,0.92}
\definecolor {gray93}              {rgb}{0.93,0.93,0.93}
\definecolor {gray94}              {rgb}{0.94,0.94,0.94}
\definecolor {gray95}              {rgb}{0.95,0.95,0.95}
\definecolor {gray97}              {rgb}{0.97,0.97,0.97}
\definecolor {gray98}              {rgb}{0.98,0.98,0.98}
\definecolor {gray99}              {rgb}{0.99,0.99,0.99}
\definecolor {darkgrey}            {rgb}{0.66,0.66,0.66}

\newcommand{\TODO}[1]{{}}

\newcommand{\ignore}[1]{}

\newcommand{\RSTODO}[1]{{\bf \textcolor{darkgreen}{{\fbox{RS TODO:} #1}}}}
\renewcommand{\RSTODO}[1]{}

\newcommand{\ignoreinshort}[1]{}
\newcommand{\ignoreinlong}[1]{{#1}}

\providecommand{\longversion}{true}
\ifthenelse{\equal{\longversion}{true}}
{%
    \renewcommand{\ignoreinshort}[1]{\textcolor{blue}{#1}}
    \newcommand{\ignoreinshortnc}[1]{{#1}}
    \renewcommand{\ignoreinlong}[1]{}
    \newcommand{\ignoreinlongnc}[1]{}
    
    \NewEnviron{ignoreinlongenv}{}
}%
{%
    \renewcommand{\ignoreinshort}[1]{}
    \newcommand{\ignoreinshortnc}[1]{}
    \renewcommand{\ignoreinlong}[1]{\textcolor{blue}{#1}}
    \newcommand{\ignoreinlongnc}[1]{{#1}}
    \NewEnviron{ignoreinshortenv}{}
    
}

\def\makenewenumerate#1#2{%
    \newcounter{cnt#1}
    \newenvironment{#1}%
    {\begin{list}{\makebox[0pt][r]{#2}}%
            {\setlength{\itemsep}{0pt}%
                \setlength{\parsep}{.2em}%
                \setlength{\leftmargin}{1.5em}%
                \setlength{\labelwidth}{.4em}%
                \usecounter{cnt#1}}}
            {\end{list}}}

\makenewenumerate{myenumerate}{\arabic{cntmyenumerate}.}

\makenewenumerate{renumerate}{\rm(\roman{cntrenumerate})}
\makenewenumerate{renumerateprime}{\rm(\roman{cntrenumerateprime}$'$)}
\makenewenumerate{renumeratesecond}{\rm(\roman{cntrenumeratesecond}$''$)}

\makenewenumerate{aenumerate}{({\it\alph{cntaenumerate}})}
\makenewenumerate{aenumerateprime}{({\it\alph{cntaenumerateprime}$'$})}
\makenewenumerate{aenumeratesecond}{({\it\alph{cntaenumeratesecond}$''$})}

\newcommand{\sref}[1]{\S{}\ref{#1}}
\newcommand{\noi}{\noindent}

\newcommand{\set}[1]{\ensuremath{\{{#1}\}}\xspace}

\renewcommand{\iff}{\ensuremath{\leftrightarrow}\xspace}
\newcommand{\defas}{\ensuremath{\stackrel{\text{\scalebox{.7}{def}}}{=}}\xspace}

\newcommand{\pos}{\phantom{\neg}}

\newcommand{\obdds}{\text{OBDD}s\xspace}
\newcommand{\obdd}{\textrm{OBDD}\xspace}
\newcommand{\obddof}[1]{\textrm{OBDD}{\ensuremath{(#1)}}\xspace}
\newcommand{\Tobdds}{\text{\T-OBDD}s\xspace}
\newcommand{\Tobdd}{\textrm{\T-OBDD}\xspace}
\newcommand{\Tobddof}[1]{\textrm{\T-OBDD}{\ensuremath{(#1)}}\xspace}
\newcommand{\sdds}{\text{SDD}s\xspace}
\newcommand{\sdd}{\textrm{SDD}\xspace}

\newcommand{\Tsdds}{\text{\T-SDD}s\xspace}
\newcommand{\Tsdd}{\textrm{\T-SDD}\xspace}

\newcommand{\bdds}{\text{DD}s\xspace}
\newcommand{\bdd}{\textrm{DD}\xspace}
\newcommand{\bddof}[1]{\textrm{DD}{\ensuremath{(#1)}}\xspace}
\newcommand{\Tbdds}{\text{\T-DD}s\xspace}
\newcommand{\Tbdd}{\textrm{\T-DD}\xspace}
\newcommand{\Tbddof}[1]{\textrm{\T-DD}{\ensuremath{(#1)}}\xspace}

\newcommand\mysout{\bgroup \markoverwith{{-}}\ULon}
\newcommand\nosout{\bgroup \markoverwith{{ }}\ULon}
\definecolor{mygray}{rgb}{0.90,0.90,0.90}
\definecolor{mywhite}{rgb}{1.00,1.00,1.00}

\newcommand{\satres}{\textsc{sat}\xspace}
\newcommand{\unsatres}{\textsc{unsat}\xspace}

\newcommand{\proptofol}{\ensuremath{{\cal B}2{\cal T}}\xspace}
\newcommand{\foltoprop}{\ensuremath{{\cal T}2{\cal B}}\xspace}
\newcommand{\btot}{\proptofol}
\newcommand{\ttob}{\foltoprop}

\newcommand{\vip}{\ensuremath{\varphi^p}\xspace}
\newcommand{\etap}{\ensuremath{\eta^p}\xspace}

\newcommand{\B}{\ensuremath{\mathcal{B}}\xspace}
\newcommand{\T}{\ensuremath{\mathcal{T}}\xspace}

\newcommand{\smt}{SMT\xspace}
\newcommand{\smtt}{\ensuremath{\text{SMT}(\T)}\xspace}

\newcommand{\euf}{\ensuremath{\mathcal{EUF}}\xspace}

\newcommand{\eq}{\ensuremath{\mathcal{E}}\xspace}
\newcommand{\dl}{\ensuremath{\mathcal{DL}}\xspace}

\newcommand{\larat}{\ensuremath{\mathcal{LA}(\mathbb{Q})}\xspace}
\newcommand{\laint}{\ensuremath{\mathcal{LA}(\mathbb{Z})}\xspace}
\newcommand{\laratint}{\ensuremath{\mathcal{LA}(\mathbb{Q}\mathbb{Z})}\xspace}

\renewcommand{\larat}{\ensuremath{\mathcal{LRA}}\xspace}
\renewcommand{\laint}{\ensuremath{\mathcal{LIA}}\xspace}
\renewcommand{\laratint}{\ensuremath{\mathcal{LA}}\xspace}

\newcommand{\nla}{\ensuremath{\mathcal{NLA}}\xspace}

\newcommand{\Tmodels}{\models_{\T}}

\newcommand{\pmodels}{\models_p}

\newcommand{\Tlemmas}{\T-lemmas\xspace}

\newcommand{\mathsat}{\textsc{MathSAT}\xspace}

\newcommand{\mathsatfive}{\textsc{MathSAT5}\xspace}

\renewcommand{\TODO}[1]{\todo[inline,color=green!40]{{\small{TODO: #1}}}}

\renewcommand{\RSTODO}[1]{\todo[inline,color=green!40]{{\small{RS TODO: #1}}}}

\newcommand{\allA}{\ensuremath{\mathbf{A}}\xspace}
\newcommand{\allalpha}{\ensuremath{\boldsymbol{\alpha}}\xspace}
\newcommand{\allB}{\ensuremath{\mathbf{B}}\xspace}
\newcommand{\allx}{\ensuremath{\mathbf{x}}\xspace}

\newcommand{\vi}{\ensuremath{\varphi}}
\newcommand{\viprime}{\ensuremath{\varphi'}}

\renewcommand{\B}{\ensuremath{\mathbb{B}}\xspace}

\newcommand{\tvpi}{\ensuremath{\mathcal{TVPI}}\xspace}

\newcommand{\betaprime}{\ensuremath{\beta'}}
\newcommand{\Bprime}{\ensuremath{B'}}
\newcommand{\allsym}[1]{\ensuremath{{\underline{\boldsymbol#1}}}}
\renewcommand{\allsym}[1]{\ensuremath{{\boldsymbol#1}}}
\renewcommand{\allA}{\allsym{A}}
\renewcommand{\allB}{\allsym{B}}
\newcommand{\allBprime}{\allsym{\Bprime{}}}
\renewcommand{\allx}{\allsym{x}}
\renewcommand{\allalpha}{\allsym{\alpha}}
\newcommand{\allbeta}{\allsym{\beta}}
\newcommand{\allbetaprime}{\allsym{\betaprime}}
\newcommand{\allvi}{\allsym{\vi}}

\newcommand{\via}[1]{\ensuremath{\varphi_{#1}[\allalpha]}}
\newcommand{\etaa}[1]{\ensuremath{\eta_{#1}[\allalpha]}}
\newcommand{\rhoa}[1]{\ensuremath{\rho_{#1}[\allalpha]}}
\newcommand{\rhoap}[1]{\ensuremath{\rho_{#1}^p[\allA]}}
\newcommand{\Ca}[1]{\ensuremath{C_{#1}[\allalpha]}}
\renewcommand{\Cap}[1]{\ensuremath{C_{#1}^p[\allA]}}
\newcommand{\viab}[1]{\ensuremath{\varphi_{#1}[\allalpha,\allbeta]}}
\newcommand{\viabb}[1]{\ensuremath{\varphi_{#1}[\allalpha,\allbeta,\allbetaprime]}}

\newcommand{\Cab}[1]{\ensuremath{C_{#1}[\allalpha,\allbeta]}}
\newcommand{\viap}[1]{\ensuremath{\varphi_{#1}^{p}[\allA]}}
\newcommand{\Cabp}[1]{\ensuremath{C_{#1}^{p}[\allA,\allB]}}

\newcommand{\etaap}[1]{\ensuremath{\eta_{#1}^{p}[\allA]}}

\newcommand{\viaprime}[1]{\ensuremath{\varphi_{#1}'[\allalpha]}}
\newcommand{\etaaprime}[1]{\ensuremath{\eta_{#1}'[\allalpha]}}

\newcommand{\viabprime}[1]{\ensuremath{\varphi_{#1}'[\allalpha,\allbetaprime]}}
\newcommand{\viabbprime}[1]{\ensuremath{\varphi_{#1}'[\allalpha,\allbeta,\allbetaprime]}}

\newcommand{\Cabprime}[1]{\ensuremath{C_{#1}'[\allalpha,\allbetaprime]}}
\newcommand{\viaprimep}[1]{\ensuremath{\varphi_{#1}'^{p}[\allA]}}
\newcommand{\Cabprimep}[1]{\ensuremath{C_{#1}'^{p}[\allA,\allBprime]}}

\newcommand{\etaaprimep}[1]{\ensuremath{\eta_{#1}'^{p}[\allA]}}

\newcommand{\vistarap}[1]{\ensuremath{\varphi_{#1}^{*p}[\allA]}}

\newcommand{\bequiv}{\ensuremath{\equiv_{\B}}}

\newcommand{\Tequiv}{\ensuremath{\equiv_{\T}}}
\newcommand{\inputcl}{\ensuremath{\vi}}
\newcommand{\outputcl}{\ensuremath{\mathcal{DD}}}
\newcommand{\theorycl}{\ensuremath{\{C_{1},\ldots,C_{K}\}}}
\newcommand{\theorycla}{\ensuremath{\{\Ca{1},\ldots,\Ca{K}\}}}

\newcommand{\theoryass}{\ensuremath{\{\eta_1,\ldots,\eta_N\}}}
\newcommand{\inputboolcl}{\ensuremath{\exists\allB.\foltoprop(\{\vi\wedge
    \bigwedge_{l=1}^K C_l\})}}
\newcommand{\outputboolcl}{\ensuremath{\foltoprop(\mathcal{DD})}}
\renewcommand{\pmodels}{\models_{\B}}

\newcommand{\theoryclab}{\ensuremath{\{\Cab{1},\ldots,\Cab{K}\}}}

\newcommand{\green}[1]{\textcolor{darkgreen}{#1}}

\newcommand{\cone}{\ensuremath{(x\le 0)}}
\newcommand{\ctwo}{\ensuremath{(x\ge 1)}}
\newcommand{\cthree}{\ensuremath{(x\le 2)}}
\newcommand{\Azero}{\ensuremath{A_0}}
\newcommand{\Aone}{\ensuremath{A_1}}
\newcommand{\Atwo}{\ensuremath{A_2}}
\newcommand{\Athree}{\ensuremath{A_3}}

\newcommand{\azero}{\ensuremath{(y \le 0)}\xspace}
\newcommand{\aone}{\ensuremath{(x \le 0)}\xspace}
\newcommand{\atwo}{\ensuremath{(x = 1)}\xspace}
\newcommand{\bone}{\ensuremath{(x \le 1)}\xspace}
\newcommand{\btwo}{\ensuremath{(x \ge 1)}\xspace}

\newcommand{\criterion}[1]{\ensuremath{\Lambda}(#1)}
\newcommand{\CTTA}[1]{\ensuremath{H}(#1)} 
\newcommand{\ITTA}[1]{\ensuremath{P}(#1)} 
\renewcommand{\CTTA}[1]{\ensuremath{H_{\allalpha}(#1)}} 
\renewcommand{\ITTA}[1]{\ensuremath{P_{\allalpha}(#1)}} 
\newcommand{\TLEMMAS}[1]{\ensuremath{Cl_{\allalpha}(#1)}}
\newcommand{\ETLEMMAS}[1]{\ensuremath{Cl_{\allalpha,\allbeta}(#1)}}
\newcommand{\ETLEMMASPRIME}[1]{\ensuremath{Cl_{\allalpha,\allbetaprime}(#1)}}
\newcommand{\EDEFS}[1]{\ensuremath{\mbox{{\em Defs}}_{\allalpha,\allbeta}(#1)}}

\begin{document}


\begin{frontmatter}


\paperid{2238} 


\title{Canonical Decision Diagrams Modulo Theories}


\author[A]{\fnms{Massimo}~\snm{Michelutti}\orcid{0009-0002-3490-2910}}

\author[A]{\fnms{Gabriele}~\snm{Masina}\orcid{0000-0001-8842-4913}}

\author[A]{\fnms{Giuseppe}~\snm{Spallitta}\orcid{0000-0002-4321-4995}}

\author[A]{\fnms{Roberto}~\snm{Sebastiani}\orcid{0000-0002-0989-6101}\thanks{Corresponding Author. Email: roberto.sebastiani@unitn.it.}}

\address[A]{DISI, University of Trento, Italy}


\begin{abstract}
Decision diagrams (\bdds{}) are powerful tools to represent
effectively propositional formulas, which are largely used in many
domains, in particular in formal verification and in knowledge
compilation. Some forms of \bdds{} (e.g., \obdds{}, \sdds{}) are {\em canonical}, that
is, 
(under given conditions on the atom list) they univocally represent equivalence
classes of formulas.
Given 
the limited expressiveness of propositional
logic, a few attempts to leverage \bdds{} to SMT level have been presented in
the literature. Unfortunately, these techniques still suffer from some
limitations:
most procedures are theory-specific;
some produce theory \bdds{} (\Tbdds) which
do not univocally represent \T-valid formulas or \T-inconsistent formulas;
none of these techniques provably produces {\em theory-canonical} \Tbdds, which
(under given conditions on the \T-atom list) 
univocally represent \T-equivalence classes of formulas.
Also, these procedures are not easy to implement, and
very few implementations are actually available.


In this paper, we present a novel very-general technique to leverage \bdds{} to
SMT level, which has several advantages: it is very easy to implement
on top of an AllSMT solver and a \bdd{} package, which are used as black boxes;
it works for every form of \bdds{} and every theory, or combination
thereof, supported by the AllSMT solver; {\em it produces
theory-canonical \Tbdds} if the propositional \bdd{}
is canonical.
We have implemented a prototype tool for both \Tobdds{} and \Tsdds{} on
top of \obdd{} and \sdd{} packages and the MathSAT SMT
solver.
Some preliminary empirical evaluation supports the
effectiveness of the approach. 

\end{abstract}

\end{frontmatter}


\section{Introduction}%
\label{sec:intro}

In the field of Knowledge Compilation (KC), the aim is to transform a given knowledge base, often represented as a Boolean formula, into a more suitable form that facilitates efficient query answering. This involves shifting the bulk of computational effort to the offline compilation phase, thereby optimizing the efficiency of the online query-answering phase.
Many representations are subsets of Negation Normal Form (NNF), and in particular of decomposable, deterministic NNF (d-DNNF)~\cite{darwicheKnowledgeCompilationMap2002}.
Among these, decision diagrams (\bdds{}) such as Ordered Binary Decision Diagrams (\obdds{})~\cite{bryantGraphBasedAlgorithmsBoolean1986} and Sentential Decision Diagrams (\sdds{})~\cite{darwicheSDDNewCanonical2011} are well-established and widely adopted representations in KC.\@ They offer efficient querying and manipulation of Boolean functions and serve as foundational elements in numerous tools across various domains, including planning~\cite{huangCombiningKnowledgeCompilation2006}, probabilistic inference~\cite{broeckCompletenessFirstOrderKnowledge2011,vandenbroeckLiftedInferenceLearning2013}, probabilistic reasoning~\cite{chaviraProbabilisticInferenceWeighted2008,fierensInferenceProbabilisticLogic2011}, and formal verification~\cite{burchSymbolicModelChecking1992}. 
Central to KC is the notion of \textit{canonicity}, where two equivalent Boolean formulas yield identical decision diagrams. Under specific conditions, both \obdds{} and \sdds{} can achieve canonicity.

The literature on KC, decision diagrams, and canonicity for Boolean
formulas is extensive. However, there is a notable scarcity of
literature addressing scenarios where formulas contain first-order
logic theories such as difference logic (\dl), two variables per
inequality~(\tvpi), linear and non-linear arithmetic (\larat{} and \nla), and equalities (\eq) with uninterpreted functions (\euf), which requires leveraging decision diagrams for Satisfiability Modulo Theories (SMT). 

\paragraph{Related Work.}
  Most of the literature has focused on theory-aware \obdds.
  The majority of the works are theory-specific, in particular focusing on \eq~\cite{frisogrooteEquationalBinaryDecision2000,goelBDDBasedProcedures1998,goelBDDBasedProcedures2003,bryantBooleanSatisfiabilityTransitivity2002}, \euf{}~\cite{vandepolBDDRepresentationLogicEquality2005,badbanAlgorithmVerifyFormulas2004,badbanZeroSuccessorEquality2005},
  and fragments of arithmetic, such as \dl~\cite{mollerDifferenceDecisionDiagrams1999}, \tvpi~\cite{chakiDecisionDiagramsLinear2009}, and \nla~\cite{chanCombiningConstraintSolving1997}. Some general approaches have been proposed to support arbitrary theories~\cite{deharbeLightweightTheoremProving2003,fontaineUsingBDDsCombinations2002,cavadaComputingPredicateAbstractions2007,cimattiTighterIntegrationBDDs2010}.
  To the best of our knowledge, the only tentative to extend \sdds{}
  to support first-order theories are
  XSDDs~\cite{dosmartiresExactApproximateWeighted2019,kolbHowExploitStructure2020}
  which support \larat{}.

  
  From the practical point of view, most of the techniques are hard to
  implement since they require modifying the internals of some SMT
  solver or \bdd{} package, or both. Indeed, all of them, with the only exception of LDDs~\cite{chakiDecisionDiagramsLinear2009}, do not have a public implementation, or are implemented within other tools, making them not directly usable and comparable to our approach.
  From the theoretical point of view,
  some techniques allow for theory-inconsistent paths (e.g., LDDs~\cite{chakiDecisionDiagramsLinear2009} and XSDDs~\cite{dosmartiresExactApproximateWeighted2019,kolbHowExploitStructure2020}), while others only guarantee theory-semicanonicity, i.e., they map all theory-valid and all theory-inconsistent formulas to the same DD (e.g., DDDs~\cite{mollerDifferenceDecisionDiagrams1999}). Notably, none of them has been proven to be theory-canonical.

  An extensive and detailed analysis of all these techniques is
  available in~\Cref{sec:apdx:rw}.

\paragraph{Contributions.}
In this paper, we investigate the problem of leveraging Boolean
decision diagrams (\bdds{}) to SMT level (\Tbdds{}). We present a
general 
formal framework for \Tbdds{}.
%
Then, we introduce a novel and highly versatile technique for
extending decision diagrams to the realm of SMT, which 
operates as follows: we perform a total enumeration of the truth
assignments satisfying the input SMT
formula (AllSMT)~\cite{lahiriSMTTechniquesFast2006}, extracting a set of theory lemmas, i.e., \T-valid clauses,
that rules out all
\T-inconsistent truth assignments. These \T-lemmas are then
conjoined to the original SMT problem, and its Boolean abstraction is
fed to a Boolean \bdd{} compiler to generate a theory \bdd{} (\Tbdd{}).  We
formally establish how our proposed framework ensures the generation
of \T-canonical decision diagrams, provided the underlying Boolean
decision diagram is canonical.

Our technique offers several advantages. Firstly, it is
very easy to implement, relying on standard AllSMT solvers and
existing \bdd{} packages as black boxes, with no need to put the hands inside the code of the AllSMT solver and of the \bdd{}
package. This simplicity makes it
accessible to a wide range of users, regardless of their expertise
level in SMT solving and DD compiling. Additionally, our technique is theory-agnostic,
accommodating any theory or combination thereof supported by the
AllSMT solver, and \bdd{}-agnostic, since it potentially works with
any form of \bdd{}. Remarkably, if the underlying \bdd{} is canonical, 
{\em it produces theory-canonical \Tbdd{}s}, ensuring that two \T-equivalent
formulas under the same set of theory atoms share the same \Tbdd{}.
Also, it is the first implementation that can be used
for \#SMT~\cite{phanModelCountingModulo2015}, because the \Tbdd{}
represents only theory-consistent truth assignments. 
Finally, our approach distinguishes itself from eager SMT encodings, which are notably quite expensive, by generating only a subset of \T-lemmas necessary to rule out \T-inconsistent assignments. 

We have implemented a prototype of \Tobdd{} and \Tsdd{} compiler based on our algorithm,
using the \mathsat{} AllSMT solver~\cite{mathsat5_tacas13} along with state-of-the-art packages
for \obdds{} and \sdds{}. A preliminary empirical evaluation demonstrates the
effectiveness of our approach in producing \T-canonical \Tobdd{}s and
\Tsdd{}s for several theories.


\section{Background}%
\label{sec:background}
\paragraph{Notation \& Terminology.}
We assume the reader is familiar with the basic syntax, semantics, and results of propositional and first-order logics.
We adopt the following terminology and notation.

Satisfiability Modulo Theories (SMT)~\cite{barrettSatisfiabilityModuloTheories2021} extends SAT to the context of first-order formulas modulo some background theory \T, which provides an intended interpretation for constant, function, and predicate symbols. We restrict to quantifier-free formulas. 
A \T-formula is a combination of theory-specific atoms (\T-atoms) via Boolean connectives.
For instance, \larat-atoms are linear (in)equalities over
rational variables.
We say that a formula is \T-satisfiable (or \T-consistent) if it is satisfiable in a model of \T{}; otherwise, we say that it is \T-unsatisfiable (or \T-inconsistent).
(For instance, $((x-y\le 3)\vee (x-y\ge 4))$ is \larat-satisfiable, whereas $((x-y\le 3)\wedge (x-y\ge 4))$ is not.)

%

%

%
\foltoprop{} is a bijective function (``theory to Boolean''),
called {\em Boolean (or propositional) abstraction},
which maps Boolean atoms into themselves,
 \T{}-atoms into fresh Boolean variables, 
and is homomorphic wrt.\ Boolean operators and set inclusion.
The function \proptofol{} (``Boolean to theory''), called {\em
  refinement},  is the inverse of \foltoprop. 
%
  (For instance
$\foltoprop(\{((x-y\le 3)\vee (x=z)) \}) =
\{(A_1\vee A_2) \}$, $A_1$ and $A_2$ being fresh
Boolean variables, and $\proptofol(\{\neg A_1, A_2\})=
\{\neg (x-y\le 3), (x=z)\}$.)

The symbols
$\allalpha\defas\set{\alpha_i}_i$,
$\allbeta\defas\set{\beta_i}_i$ denote ground \T-atoms on \T-variables
$\allx\defas\set{x_i}_i$.
The symbols
$\allA\defas\set{A_i}_i$,
$\allB\defas\set{B_i}_i$ denote Boolean atoms, and typically denote also the
Boolean abstraction of the \T-atoms in \allalpha{}, \allbeta{}
respectively.
(Notice that a Boolean atom is also a \T-atom, which is mapped into itself
by \ttob{}.)
We represent truth assignments as conjunctions of literals.
We
denote by
$2^{\allalpha}$ the set of all total truth assignments on \allalpha.
The symbols
$\vi$, $\psi$ denote \T{}-formulas, and 
$\mu$, $\eta$, $\rho$ denote conjunctions of \T{}-literals;
 $\vi^p$, $\psi^p$  denote Boolean formulas,
 $\mu^p$,  $\eta^p$, $\rho^p$  
denote conjunctions of Boolean literals (i.e., truth assignments) 
and we use them as synonyms for the Boolean abstraction of
$\vi$, $\psi$, $\eta$,  and $\rho$ respectively,  and vice versa 
(e.g., $\vi^p$ denotes $\foltoprop(\vi)$,
$\eta$ denotes $\proptofol(\etap)$). 
If $\foltoprop(\eta) \models \foltoprop(\vi)$, then we say that $\eta$
\emph{propositionally satisfies} $\vi$, written
$\eta\pmodels\vi$. 
The notion of propositional 
entailment and validity follow straightforwardly.
When both $\vi\pmodels\psi$ and $\psi\pmodels\vi$, we say that
$\vi$ and $\psi$ are {\em propositionally equivalent}, written 
``$\vi\bequiv\psi$''.
%
``$\Tmodels$'' denotes entailment in $\T{}$ (e.g. $(x \geq 2) \models_{\larat} (x \geq 1)$). Notice that if $\eta\pmodels\vi$ then $\eta\Tmodels\vi$, but not vice versa. (E.g., $(x \geq 2) \not\pmodels (x \geq 1)$.)
%
When both $\vi\Tmodels\psi$ and $\psi\Tmodels\vi$, we say that
$\vi$ and $\psi$ are {\em \T-equivalent}, written 
``$\vi\Tequiv\psi$''.
 Notice that if $\eta\bequiv\vi$ then
 $\eta\Tequiv\vi$, but not vice versa.
We call a {\em \T-lemma} any \T-valid clause.


We denote by \via{} the fact that  \allalpha{} is a superset of the
set of \T-atoms occurring in $\vi$ whose truth assignments we are interested
in. The fact that it is a superset is sometimes necessary for
comparing formulas with different sets of \T-atoms:
$\via{}$ and $\viprime[\allalpha{}']$ can be compared only if they are both
considered as formulas on $\allalpha{}\cup\allalpha{}'$.
(E.g., in order to check that $(A_1\vee A_2)\wedge (A_1\vee\neg A_2)$ and
$(A_1\vee A_3)\wedge (A_1\vee\neg A_3)$ are equivalent, we need
considering them as formulas 
on \set{A_1,A_2,A_3}.)

\paragraph*{Decision Diagrams.}
Knowledge compilation is the process of transforming a formula into a representation that is more suitable for answering queries~\cite{darwicheKnowledgeCompilationMap2002}.
Many known representations are subsets of Negation Normal Form (NNF), which requires formulas to be represented by Directed Acyclic Graphs (DAGs) where internal nodes are labelled with $\land$ or $\lor$, and leaves are labelled with literals $A,\neg A$, or constants $\top,\bot$. Other languages are defined as special cases of NNF~\cite{darwicheKnowledgeCompilationMap2002}.
In particular, Decision Diagrams (\bdds{}) like \obdds{} and \sdds{} are
popular compilation languages. 

Ordered Binary Decision Diagrams (\obdds)~\cite{bryantGraphBasedAlgorithmsBoolean1986} are NNFs where the root node is a decision node and a total order ``$<$'' on the atoms is imposed.
A decision node is either a constant $\top, \bot$, or a $\vee$-node having the form 
$(A \wedge \vi) \vee (\neg A \wedge \viprime)$, where $A$ is an atom, and $\vi,\viprime$ are decision nodes. 
In every path from the root to a leaf, each atom is tested at most only once, following the order ``$<$''.
\Cref{fig:bdd-sdd} (left) shows a graphical representation of an
\obdd, where each decision node is graphically represented as a node
labelled with the atom $A$ being tested; a solid and a dashed edge
connect it to the nodes of \vi{} and \viprime{}, representing the
cases in which $A$ is true or false, respectively.\footnote{In all examples we use \obdds{} only because
  it is 
  eye-catching to detect the partial assignments which verify or
  falsify the formula.}
%
\obdds{} allow performing many operations in polynomial time, such as
performing Boolean combinations of \obdds{}, or checking for
(un)satisfiability or validity.

\sdds~\cite{darwicheSDDNewCanonical2011} are a generalization of \obdds{}, in which decisions are not binary and are made on sentences instead of atoms. Formally, an \sdd{} is an NNF that satisfies the properties of \emph{structured decomposability} and \emph{strong determinism}. A v-tree $v$ for atoms $\allA$ is a full binary tree whose leaves are in one-to-one correspondence with the atoms in $\allA$. We denote with $v_l$ and $v_r$ the left and right subtrees of $v$.
An \sdd{} that respects $v$ is either: a constant $\top, \bot$; a literal $A, \neg A$ if $v$ is a leaf labelled with $A$; a decomposition $\bigvee_{i=1}^n (\vi_i \wedge \psi_i)$ if $v$ is internal,
$\vi_1,\ldots,\vi_n$ are \sdds{} that respect subtrees of $v_l$, $\psi_1,\ldots,\psi_n$ are \sdds{} that respect subtrees of $v_r$, and $\allvi\defas\set{\vi_1,\ldots,\vi_n}$ is a partition. \allvi{} is called a partition if each $\vi_i$ is consistent, every pair $\vi_i, \vi_j$ for $i\neq j$ are mutually exclusive, and the disjunction of all $\vi_i$s is valid.
$\vi_i$s are called \emph{primes}, and $\psi_i$s are called \emph{subs}. A pair $\vi_i,\psi_i$ is called an \emph{element}.
\Cref{fig:bdd-sdd} (right) shows a graphical representation of an \sdd. Decomposition nodes are represented as circles, with an outgoing edge for each element. Elements are represented as paired boxes, where the left and right boxes represent the prime and the sub, respectively.
\sdds{} maintain many of the properties of \obdds{}, with the advantage of being exponentially more succinct~\cite{bovaSDDsAreExponentially2016}. 

These forms of \bdds{} are \emph{canonical modulo some canonicity
  condition} \criterion{\allA} on the Boolean atoms \allA:\@
under the assumption that the \bdds{} are built according to the same canonicity
condition \criterion{\allA}, 
then each formula $\vi^p[\allA]$ has a unique \bdd{} representation $\bddof{\vi^p[\allA]}$, and 
$\bddof{\vi^p[\allA]}=\bddof{\viprime^{p}[\allA]}$ if and only if
$\vi^p[\allA]\equiv\viprime^{p}[\allA]$ and hence
if and only if
$\bddof{\vi^p[\allA]}\equiv\bddof{\viprime^{p}[\allA]}$.
(E.g., for
\obdds, \criterion{\allA} is given total order on \allA{}~\cite{bryantGraphBasedAlgorithmsBoolean1986}; for \sdds{}
\criterion{\allA} is the structure
induced by a given v-tree{}~\cite{darwicheSDDNewCanonical2011}.)
Canonicity allows easily checking if a formula is a tautology or a contradiction, and if two formulas are equivalent. Also, it allows storing equivalent subformulas only once.

\begin{figure}[t]
  \centering
  \begin{tikzpicture}[scale=0.8, transform shape, every node/.style={rectangle, draw, minimum size=6mm}, circle_node/.style={circle, draw, minimum size=6mm, rounded corners=4pt}]
    \node (A) at (0,0) {$A_1$};
    \node (B) at (1,-1) {$A_2$};
    \node (C) at (-1,-2) {$A_3$};
    \node (D) at (0,-3) {$A_4$};
    \node[circle_node] (F) at (-1,-4) {$\bot$};
    \node[circle_node] (T) at (1,-4) {$\top$};
    \draw (A) -- (B);
    \draw[dashed] (A) -- (C);
    \draw[dashed] (B) -- (C);
    \draw (B) -- (T);
    \draw (C) -- (D);
    \draw (D) -- (T);
    \draw[dashed] (C) -- (F);
    \draw[dashed] (D) -- (F);
\end{tikzpicture}
\hspace{1cm}
  \begin{tikzpicture}[scale=0.8, transform shape, 
    every node/.style={rectangle, draw, minimum size=6mm},
    circle_node/.style={circle, draw, minimum size=6mm, rounded corners=4pt}]

    \node[circle_node] (circle1) at (0.2,0) {};
    
    \node (A1) at (-1.5,-1) {$A_1$};
    \node[right=0mm of A1] (dot-A1) {$\bullet$};
    \node (A2) at (1.5,-1) {$\neg A_1$};
    \node[right=0mm of A2] (dot-A2) {$\bullet$};
    
    \node[circle_node] (circle2) at (-0.9,-2) {};
    \node (B1) at (-3,-3) {$A_2$};
    \node[right=0mm of B1] (dot-B1) {$\top$};
    \node (B2) at (-1,-3) {$A_2$};
    \node[right=0mm of B2] (dot-B2) {$\bullet$};

    \node[circle_node] (circle3) at (1,-3.5) {};
    \node (C1) at (-0.5,-4.5) {$A_3$};
    \node[right=0mm of C1] (dot-C1) {$A_4$};
    \node (C2) at (1.5,-4.5) {$\neg A_3$};
    \node[right=0mm of C2] (dot-C2) {$\bot$};
    \draw[->] (circle1) -- (A1.north east);
    \draw[->] (circle1) -- (A2.north east);
    \draw[->] (dot-A1.center) -- (circle2);
    \draw[->] (dot-A2.center) -- (circle3);
    \draw[->] (circle2) -- (dot-B1.north west);
    \draw[->] (circle2) -- (B2.north east);
    \draw[->] (circle3) -- (dot-C1.north west);
    \draw[->] (circle3) -- (C2.north east);
    \draw[->] (dot-B2.center) -- (circle3);
\end{tikzpicture}
    \vspace{.4cm}
  \caption{\obdd{} (left) and \sdd{} (right) for \\ $\vi = (A_1 \wedge A_2) \vee (A_2 \wedge A_3) \vee (A_3 \wedge A_4)$. \\ \\ }%
  \label{fig:bdd-sdd}
\end{figure}

\section{A Formal Framework for \Tbdds}%
\label{sec:theory}
  In this section, we introduce the theoretical results that will be
  used in the rest of the paper. For the sake of compactness, all the
  proofs of the theorems are deferred to~\Cref{sec:apdx:proofs}.  

Given a set \allalpha{} of \T-atoms and a \T-formula \via{}, we denote by
$\CTTA{\vi{}}\defas\set{\etaa{i}}_{i}$ and 
$\ITTA{\vi{}}\defas\set{\rhoa{j}}_{j}$ respectively the set of all
{\em \T-consistent} and that of all {\em \T-inconsistent} total truth
assignments on the set $\allalpha{}$ of \T-atoms which
propositionally satisfy $\vi$, i.e., s.t.
\begin{eqnarray}
  \label{eq:decomposition}
  \via{} \bequiv \bigvee_{\etaa{i}\in\CTTA{\vi}}\etaa{i} \vee \bigvee_{\rhoa{j}\in\ITTA{\vi}}\rhoa{j}.
\end{eqnarray}
%
%
The following facts are straightforward consequences of the definition
of $\CTTA{\vi{}}$ and $\ITTA{\vi{}}$.

\begin{proposition}
  Given two \T-formulas \via{} and \viaprime{}, we have: 
  \begin{aenumerate}%
  \label{prop:assignmentsets}
  \item $\CTTA{\vi{}}$, $\ITTA{\vi{}}$, $\CTTA{\neg\vi{}}$,
    $\ITTA{\neg\vi{}}$ are pairwise disjoint;
  \item $\CTTA{\vi{}} \cup \ITTA{\vi{}} \cup \CTTA{\neg\vi{}} \cup \ITTA{\neg\vi{}}=2^{\allalpha}$; 
  \item $\via{}\bequiv\viaprime{}$ iff
    $\CTTA{\vi{}}=\CTTA{\viprime{}}$ and $\ITTA{\vi{}}=\ITTA{\viprime{}}$;
  \item $\via{}\Tequiv\viaprime{}$ iff $\CTTA{\vi{}}=\CTTA{\viprime{}}$.
  \end{aenumerate}
\end{proposition}

\begin{example}%
\label{ex:proposition1}
  Let $\allalpha{}\defas\set{\aone,\atwo}$,
  $\via{1}\defas\aone\vee\atwo$ and
  $\via{2}\defas \neg\aone \iff \atwo$, so that
  $\viap{1}\defas \Aone \vee \Atwo$ and
  $\viap{2}\defas \neg\Aone \iff \Atwo$.  
It is easy to see that
$\via{1}\not\bequiv\via{2}$ and 
$\via{1}\equiv_{\larat}\via{2}$. 
Then $\CTTA{\via{1}}=\CTTA{\via{2}}=\set{\eta_1,\eta_2}=
\set{\aone\wedge\neg\atwo,\neg\aone\wedge\atwo}$, whereas 
$\ITTA{\via{1}}=\set{\aone\wedge\atwo}$ and
$\ITTA{\via{2}}=\emptyset$.

\end{example}



\subsection{Canonicity for \smtt{} formulas}



\noindent
%

\begin{definition}%
  \label{def:tbdd}
  Given a set \allalpha{} of \T-atoms and its
  Boolean abstraction $\allA\defas\ttob(\allalpha{})$, some 
  \T-formula \via{},
  and some form of \bdds{}  with  canonicity condition
  \criterion{\allA} (if any), 
  we call ``\Tbddof{\via{}}''
  with canonicity condition
  \criterion{\allalpha} an \smtt{} formula $\Psi[\allalpha]$
  such that $\Psi[\allalpha]\Tequiv\via{}$ and its Boolean abstraction
  $\Psi^p[\allA]$ is a
  \bdd{}.
\end{definition}
%

E.g. ``\Tobdds'' and ``\Tsdds'' denote \smtt{} extensions of \obdds{} and
\sdds{} respectively.


\begin{figure}[t]
\centering
    \begin{tikzpicture}[scale=0.9, transform shape, every node/.style={rectangle, draw, minimum size=6mm}, circle_node/.style={circle, draw, minimum size=6mm, rounded corners=4pt}]
    \node (A1) at (0,0) {$A_1$};
    \node (A2) at (-1,-1.5) {$A_2$};
    \node[circle_node] (F) at (-1,-3) {$\bot$};
    \node[circle_node] (T) at (1,-3) {$\top$};
    \draw (A1) -- (T);
    \draw[dashed] (A1) -- (A2);
    \draw (A2) -- (T);
    \draw[dashed] (A2) -- (F);
\end{tikzpicture}
\hspace{1cm}
\begin{tikzpicture}[scale=0.9, transform shape, every node/.style={rectangle, draw, minimum size=6mm}, circle_node/.style={circle, draw, minimum size=6mm, rounded corners=4pt}]
    \node (A1) at (0,0) {$x \leq 0$};
    \node (A2) at (-1,-1.5) {$x = 1$};
    \node[circle_node] (F) at (-1,-3) {$\bot$};
    \node[circle_node] (T) at (1,-3) {$\top$};
    \draw (A1) -- (T);
    \draw[dashed] (A1) -- (A2);
    \draw (A2) -- (T);
    \draw[dashed] (A2) -- (F);
\end{tikzpicture}
\\
\begin{tikzpicture}[scale=0.9, transform shape, every node/.style={rectangle, draw, minimum size=6mm}, circle_node/.style={circle, draw, minimum size=6mm, rounded corners=4pt}]
    \node (A1) at (0,0) {$A_1$};
    \node (A2_l) at (-1,-1.5) {$A_2$};
    \node (A2_r) at (1,-1.5) {$A_2$};
    \node[circle_node] (F) at (-1,-3) {$\bot$};
    \node[circle_node] (T) at (1,-3) {$\top$};
    \draw (A1) -- (A2_r);
    \draw[dashed] (A1) -- (A2_l);
    \draw (A2_l) -- (T);
    \draw[dashed] (A2_l) -- (F);
    \draw (A2_r) -- (T);
    \draw[dashed] (A2_r) -- (F);
\end{tikzpicture}
\hspace{1cm}
\begin{tikzpicture}[scale=0.9, transform shape, every node/.style={rectangle, draw, minimum size=6mm}, circle_node/.style={circle, draw, minimum size=6mm, rounded corners=4pt}]
    \node (A1) at (0,0) {$x \leq 0$};
    \node (A2_l) at (-1,-1.5) {$x = 1$};
    \node (A2_r) at (1,-1.5) {$x = 1$};
    \node[circle_node] (F) at (-1,-3) {$\bot$};
    \node[circle_node] (T) at (1,-3) {$\top$};
    \draw (A1) -- (A2_r);
    \draw[dashed] (A1) -- (A2_l);
    \draw (A2_l) -- (T);
    \draw[dashed] (A2_l) -- (F);
    \draw (A2_r) -- (T);
    \draw[dashed] (A2_r) -- (F);
\end{tikzpicture}
\vspace{.4cm}
    \caption{
    Top: \obdd{} for  $\Aone \vee \Atwo$ and its refinement for $\aone \vee \atwo$. \newline
    Bottom:
  \obdd{} for  $\neg\Aone \iff \Atwo$ and its refinement for $\neg\aone \iff \atwo$. \\ \\}%
  \label{fig:noncanonical}
\end{figure}

\begin{theorem}%
    \label{teo:basicprop}
    Consider some form of \Tbdd{} such that its Boolean abstraction
    \bdd{} is canonical. Then
    $\Tbddof{\via{}}=\Tbddof{\viaprime{}}$ if and only if
    $\Tbddof{\via{}}\bequiv\Tbddof{\viaprime{}}$.
\end{theorem}

There are potentially many possible ways by which \bdds{} can be extended into \Tbdds{},
depending mainly on how the \T-consistency of branches and subformulas
is handled.
A straightforward way would be to define them as the refinement of the
\bdd{} of the Boolean abstraction, i.e.\ $\Tbddof{\via{}}\defas\btot(\bddof{\viap{}})$, without pruning
\T-inconsistent branches. Such \Tbdds{}, however,
would be neither \T-canonical nor \T-semicanonical, as defined below.

\begin{definition}%
\label{def:canonicity}
Let \allalpha{} be a set of \T-atoms, and let \criterion{\allalpha}
 be
 some canonicity condition.\\
We say that a form of  \Tbdd{} is {\bf \B-canonical}
wrt.\ \criterion{\allalpha} iff, for every pair of formulas \via{} and
\viaprime{}, $\Tbddof{\via{}}=\Tbddof{\viaprime{}}$ if 
$\via{}\bequiv\viaprime{}$.
\\
We say that a form of  \Tbdd{} is {\bf \T-canonical}
wrt. \criterion{\allalpha} iff, for every pair of formulas \via{} and
\viaprime{}, $\Tbddof{\via{}}=\Tbddof{\viaprime{}}$ if and only if
$\via{}\Tequiv\viaprime{}$.
\\
We say that a form of  \Tbdd{} is {\bf \T-semicanonical}
wrt.\ \criterion{\allalpha} iff, for every pair of formulas \via{} and
\viaprime{}, if \via{} and \viaprime{} are both \T-inconsistent
or are both \T-valid, then 
$\Tbddof{\via{}}=\Tbddof{\viaprime{}}$.
\end{definition}
\noindent
If \Tbdd{} is \T-canonical, then it is also
  \T-semicanonical, but not vice versa.
As a consequence of \Cref{teo:basicprop}, \Tbdd{} is
\B-canonical if its corresponding \bdd{} is canonical, but not vice versa.

Notice the ``if'' rather than ``if and only if'' in the definition of
{\bf \B-canonical}: it may be the case that $\Tbddof{\via{}}=\Tbddof{\viaprime{}}$ even if 
$\via{}\not\bequiv\viaprime{}$ (e.g., if $\via{}\Tequiv\viaprime{}$,
as in the case of \T-canonicity). 
%
\begin{example}
  Let \Tobdd{} be defined as $\Tobddof{\via{i}}\defas\btot(\obddof{\viap{i}})$. 
  Consider the formulas $\via{1},\via{2}$ in
  \Cref{ex:proposition1}.
  \Cref{fig:noncanonical} shows the \obdds{}
  for \viap{1} and
 \viap{2} (left) --considering as canonicity condition
 \criterion{\allA} the order \set{A_1,A_2}--  and the \Tobdds{} for
 \via{1} and 
 \via{2} (right).
 Notice that the two \Tobdds{} are different
 despite the fact that $\via{1}\Tequiv\via{2}$. Thus this form of \Tobdds{}
 is  not \T-canonical.
 \\
 Consider  the \T-valid \T-formulas $\via{3}\defas
   (\aone\vee\neg\aone)\wedge(\atwo\vee\neg\atwo)$ and $\via{4}\defas\neg\aone \vee
   \neg\atwo$, so that
   $\viap{3}= (\Aone\vee\neg\Aone)\wedge(\Atwo\vee\neg\Atwo)$ and $\viap{4}=\neg\Aone \vee
   \neg\Atwo$.
   Since \viap{3} is propositionally valid whereas \viap{4} is not, then
   \Tobddof{\via{3}} reduces to the $\top$ node whereas \Tobddof{\via{4}}
   does not. Dually, $\neg\via{3}$ and $\neg\via{4}$ are both
   \T-inconsistent,
    and \Tobddof{\neg\via{3}} reduces to the $\bot$ node whereas \Tobddof{\neg\via{4}}
   does not. Thus this form of \Tobdds{}
 is  not \T-semicanonical.
\end{example}

\begin{theorem}%
    \label{teo:canonicity-sufficient}
    Consider a form of \Tbdds{} which are  \B-canonical wrt.\ some
    canonicity condition \criterion{\allalpha}.
    Suppose that, for every \smtt{} formula
    $\via{}$,
    $\Tbddof{\via{}}\bequiv \bigvee_{\eta_i\in\CTTA{\vi{}}}\eta_i$.
    Then \Tbdd{} are \T-canonical wrt.\ \criterion{\allalpha}.
\end{theorem}
\Cref{teo:canonicity-sufficient} states a sufficient condition
to guarantee the \T-canonicity of some form of \Tbdd: it should
represent all and only \T-consistent total truth
assignments propositionally satisfying the formula.
Since typically \Tbdds{} represent {\em partial} assignments $\mu_i$, the
latter ones should not have \T-inconsistent total extensions.
%
\subsection{Canonicity via \T-lemmas}%
\label{sec:teo}

\begin{definition}%
  \label{def:ruleout-nobetas}
  We say that a set \theorycla{} of \T-lemmas {\bf rules out} a set
  $\set{\rhoa{1},\ldots,\rhoa{M}}$ of 
  \T-inconsistent total truth assignments 
  if and only if,
  for every $\rhoa{j}$ in the set, there exists a $\Ca{l}$
  s.t.\ $\rhoa{j}\pmodels \neg \Ca{l}$, that is, if and only if
  \begin{equation}
    \bigvee_{j=1}^M\rhoa{j}\wedge\bigwedge_{l=1}^K\Ca{l}\bequiv\bot.    
  \end{equation}
\end{definition}

 Given \allalpha{} and some \T-formula \via{}, we denote
 as $\TLEMMAS{\vi}$ any function which returns a set 
 \theorycla{} of \T-lemmas which rules out \ITTA{\vi}.


\begin{theorem}%
    \label{teo:teo2}
    Let \via{} be a \T-formula.
    Let $\TLEMMAS{\vi}\defas\theorycla{}$  be a set of \T-lemmas which rules out 
    \ITTA{\vi}.
    Then we have that:
    \begin{eqnarray}
        \label{eq:teo-th2-prop}
        \viap{}\  \wedge \bigwedge_{\Ca{l}\in\TLEMMAS{\vi}} \Cap{l}\
        &\equiv&\ \bigvee_{\etaa{i}\in\CTTA{\vi{}}}\etaap{i}.
    \end{eqnarray}

\end{theorem}

\begin{theorem}%
    \label{teo:canonicity-technique-noextraatoms}
    Let \allalpha{} denote arbitrary sets of \T-atoms with Boolean
    abstraction \allA{}.
    Consider some canonical form of \bdds{} on some canonicity condition
    \criterion{\allA}.
    Let \Tbdd{} with canonicity condition
    \criterion{\allalpha{}} be such that, for all sets \allalpha{}
    and for all formulas \via{}:
    \begin{eqnarray}
        \label{eq:canonicity-technique-noextraatoms}
        \Tbdd{}(\via{})&\defas& \btot{}
        (\bdd\left(\viap{}\wedge\hspace{-.4cm} \bigwedge_{\Ca{l}\in\TLEMMAS{\vi}}\hspace{-.4cm}\Cap{l}\right)).
    \end{eqnarray}
    Then the \Tbdds{} are \T-canonical.
\end{theorem}

\Cref{teo:teo2,teo:canonicity-technique-noextraatoms} 
suggests an easy
way to implement \T-canonical \Tbdds{} by using as \TLEMMAS{\vi} the
list of \Tlemmas{} produced by an \smtt{} solver during an AllSMT
run over \via{}. (We will discuss this technique in \sref{sec:tdds}.)

\subsection{Dealing with Extra \T-Atoms}

Unfortunately, things are not so simple in practice.
In order to cope with some  theories, AllSMT
solvers frequently need to introduce extra \T-atoms \allbeta{}
on-the-fly, and need to generate a set $\EDEFS{\vi}\defas\set{\Cab{l}}_l$ of some extra \T-lemmas{}, that relate the
novel atoms \allbeta{} with those occurring in the original formula
\via{}~\cite{barrettSatisfiabilityModuloTheories2021}.
%
Consequently, in this case the
list of \Tlemmas{}  produced by an \smtt{} solver during an AllSMT
run over \via{} may contain some of such \T-lemmas, and thus they cannot be used as \TLEMMAS{\vi}. 

For instance,  when the \larat-atom{} $(\sum_i a_i x_i = b)$
occurs in a \larat{}-formula,  the SMT solver may need introducing also
the \larat-atoms $(\sum_i a_i x_i\ge b)$ and $(\sum_i a_i x_i\le b)$ 
and adding some or all the \larat-lemmas encoding
$(\sum_i a_i x_i = b)\iff ( (\sum_i a_i x_i\ge b) \wedge (\sum_i a_i x_i\le b))$. 
%
\begin{remark} The number of novel theory atoms introduced by enumeration highly depends on the specific theory and on the solving procedure (for instance, \laint{} generally generates more lemmas than \larat{} due to the branch-and-bound process inherent to integer solving). For some theories (for instance \larat{}), it is possible to avoid introducing new atoms by setting some tool-specific options. Some other theories instead, and the combination of theories, require the introduction of new \T-atoms, justifying our particular attention on this aspect. 
In general, we have observed empirically that the number of new theory atoms contained in theory lemmas is limited with respect to the number of atoms in the original formula. \end{remark}
%
%
\begin{definition}%
\label{def:ruleout}  
  We say that a set \theoryclab{} of \T-lemmas on
  \allalpha{},\allbeta{} {\bf rules out} a set
  $\set{\rhoa{1},\ldots,\rhoa{M}}$ of 
  \T-inconsistent total truth assignments on \allalpha{}
  if and only if
  \begin{eqnarray}
    \label{eq:ruleout}
   \bigvee_{j=1}^M\rhoa{j}\wedge\bigwedge_{l=1}^K\Cab{l}\bequiv\bot.
  \end{eqnarray}
\end{definition}

 Given \allalpha{}, \allbeta{} and some \T-formula \via{}, we denote
 as $\ETLEMMAS{\vi}$ any function which returns a set \theoryclab{} 
 of \T-lemmas on \allalpha{}, \allbeta{} which rules out
 \ITTA{\vi}. (If $\allbeta=\emptyset$, then
 Definition~\ref{def:ruleout} reduces to
 Definition~\ref{def:ruleout-nobetas} and  $\ETLEMMAS{\vi}$ reduces to  $\TLEMMAS{\vi}$.)
 Notice that $\ETLEMMAS{\vi}$ is not unique, and it is not necessarily
 minimal: if $\Cab{}\not\in\ETLEMMAS{\vi}$ is a \T-lemma, then
 $\set{\Cab{}}\cup\ETLEMMAS{\vi}$ rules out \ITTA{\vi} as well.
 The same fact applies to \TLEMMAS{\vi} as well.

\begin{theorem}%
    \label{teo:teo-ext}
    Let \allalpha{} and \allbeta{} be sets of \T-atoms
    and let \allA{} and \allB{} denote their Boolean
    abstraction.
    Let \via{} be a \T-formula.
    Let   \ETLEMMAS{\vi{}}\defas\theoryclab{} be a set of \T-lemmas on
    \allalpha{},\allbeta{} which rules out \ITTA{\vi}.
    %
    Then we have that:
    \begin{eqnarray}
        \label{eq:teo-th3-prop}
        \viap{}  \wedge \exists\allB.\hspace{-.3cm}\bigwedge_{\Cab{l}\in\ETLEMMAS{\vi}}\hspace{-.3cm}\Cabp{l}\
        &\equiv&\hspace{-.3cm}\bigvee_{\etaa{i}\in\CTTA{\vi}}\hspace{-.3cm}\etaap{i}.
    \end{eqnarray}
\end{theorem}

\begin{example}%
  \label{ex:extraatoms}
  Consider the formula  $\via{1}\defas \aone \vee \atwo$ and its
  Boolean abstraction $\viap{1}\defas \Aone \vee \Atwo$, as in
  \Cref{ex:proposition1}.
  If we run an AllSMT solver over it we obtain the set $\set{\eta_1,\eta_2}\defas
  \set{\aone\wedge\neg\atwo,\neg\aone\wedge\atwo}$ of \T-satisfiable truth assignments, but instead
  of the \T-lemma $\neg\aone\vee\neg\atwo$, we might obtain five
  \T-lemmas:\\
  $
  C_1:\ \neg\aone\vee\neg\btwo\\
  C_2:\ \neg\aone\vee\pos\bone\\
    \!C_3:\ \neg\atwo\vee\pos\bone\\
    \!C_4:\ \neg\atwo\vee\pos\btwo\\
    \!C_5:\ \pos\atwo\vee\neg\bone\vee\neg\btwo,
  $
  \\
  because the SMT solver has introduced the extra atoms
  $\set{\beta_1,\beta_2}\defas\set{\bone,\btwo}$ and added
  the axiom
  $\atwo \iff (\bone\wedge\btwo)$, which is returned in the list of
  the \T-lemmas as $C_3,C_4,C_5$.
  \\
Now,
 if we applied the \obdd{} construction simply to
 $\viap{}\wedge\bigwedge_{i=1}^5\Cabp{i}$ and
 $\via{}\wedge\bigwedge_{i=1}^5\Cab{i}$ respectively,
  we would obtain the \obdds{} in
\Cref{fig:withaxioms}, which are much bigger than necessary. 
Instead, computing the Boolean abstraction and applying the existential
quantification on \allB{} by Shannon's expansion, we obtain:
\vspace{-.3cm}
\begin{eqnarray}
\label{eq:extraatoms}
\vip\wedge\exists B_1B_2.\bigwedge_{l=1}^5C_l^p\ \equiv\
  (\overbrace{\pos\Aone\wedge\neg\Atwo}^{\etaap{1}})\vee
(\overbrace{\neg\Aone\wedge\pos\Atwo}^{\etaap{2}}),
\end{eqnarray}
in line with~\eqref{eq:teo-th3-prop} in \Cref{teo:teo-ext}.
The resulting \obdd{} is that of \Cref{fig:noncanonical}, bottom left.
\end{example}
\begin{figure}[t]
  \begin{tikzpicture}[scale=0.8, transform shape, every node/.style={rectangle, draw, minimum size=6mm}, circle_node/.style={circle, draw, minimum size=6mm, rounded corners=4pt}]
    \node (A1) at (0,0) {$A_1$};
    
    \node (A2-1) at (-1,-1) {$A_2$};
    \node (A2-2) at (1,-1) {$A_2$};

    \node (B1-1) at (-1,-2) {$B_1$};
    \node (B1-2) at (1,-2) {$B_1$};

    \node (B2-1) at (-1,-3) {$B_2$};
    \node (B2-2) at (1,-3) {$B_2$};
    
    \node[circle_node] (F) at (-1,-4) {$\bot$};
    \node[circle_node] (T) at (1,-4) {$\top$};
    \draw (A1) -- (A2-2);
    \draw[dashed] (A1) -- (A2-1);
    \draw (A2-2) -- (F);
    \draw[dashed] (A2-2) -- (B1-2);
    \draw (A2-1) -- (B1-1);
    
    \draw (B1-2) -- (B2-2);
    \draw[dashed] (B1-2) -- (F);
    \draw (B1-1) -- (B2-1);

    \draw (B2-2) -- (F);
    \draw[dashed] (B2-2) -- (T);
    \draw (B2-1) -- (T);
    \draw[dashed] (B2-1) -- (F);

    \draw[dashed] (A2-1) .. controls +(-180:2) and +(180:2) .. (F);
    \draw[dashed] (B1-1) .. controls +(-160:1) and +(160:1) .. (F);

\end{tikzpicture}
\hspace{1cm}
 \begin{tikzpicture}[scale=0.8, transform shape, every node/.style={rectangle, draw, minimum size=6mm}, circle_node/.style={circle, draw, minimum size=6mm, rounded corners=4pt}]
    \node (A1) at (0,0) {$x \leq 0$};
    
    \node (A2-1) at (-1,-1) {$x = 1$};
    \node (A2-2) at (1,-1) {$x = 1$};

    \node (B1-1) at (-1,-2) {$x \leq 1$};
    \node (B1-2) at (1,-2) {$x \leq 1$};

    \node (B2-1) at (-1,-3) {$x \geq 1$};
    \node (B2-2) at (1,-3) {$x \geq 1$};
    
    \node[circle_node] (F) at (-1,-4) {$\bot$};
    \node[circle_node] (T) at (1,-4) {$\top$};
    \draw (A1) -- (A2-2);
    \draw[dashed] (A1) -- (A2-1);
    \draw (A2-2) -- (F);
    \draw[dashed] (A2-2) -- (B1-2);
    \draw (A2-1) -- (B1-1);
    
    \draw (B1-2) -- (B2-2);
    \draw[dashed] (B1-2) -- (F);
    \draw (B1-1) -- (B2-1);

    \draw (B2-2) -- (F);
    \draw[dashed] (B2-2) -- (T);
    \draw (B2-1) -- (T);
    \draw[dashed] (B2-1) -- (F);

    \draw[dashed] (A2-1) .. controls +(-180:2) and +(180:2) .. (F);
    \draw[dashed] (B1-1) .. controls +(-160:1) and +(160:1) .. (F);

\end{tikzpicture}
    \vspace{.4cm}
  \caption{
  \obdd{} for  $\viap{}\wedge\bigwedge_{i=1}^5\Cabp{}$ and its refinement for $\via{}\wedge\bigwedge_{i=1}^5\Cab{}$. \\ \\}%
  \label{fig:withaxioms}
\end{figure}

\begin{theorem}%
    \label{teo:canonicity}
    %
    Let \allalpha{}, \allbeta{} and \allbetaprime{} be sets of \T-atoms
    and let \allA{}, \allB{} and \allBprime{} denote their Boolean
    abstraction.
    Let \via{} and $\viaprime{}$ be \T-formulas.
    Let \ETLEMMAS{\vi{}} be a set of \T-lemmas on
    \allalpha{},\allbeta{}
    which rules out \ITTA{\vi} and
    \ETLEMMASPRIME{\viprime{}} be a set of \T-lemmas on
    \allalpha{},\allbetaprime{}
    which rules out \ITTA{\viprime}.
    Then  $\via{}\Tequiv\viaprime{}$ if and only if
    \begin{eqnarray}%
    \label{eq:teo4-th-prop}
    \viap{}  \wedge \exists\allB.\hspace{-.9cm}\bigwedge_{\Cab{l}\in\ETLEMMAS{\vi}} \hspace{-0.9cm}\Cabp{l}
    \equiv 
    \viaprimep{}  \wedge \exists\allBprime.\hspace{-1cm}\bigwedge_{\Cabprime{l}\in\ETLEMMAS{\viprime}}\hspace{-1cm}\Cabprimep{l}.
\end{eqnarray}
\end{theorem}
\noindent
As a direct consequence of \Cref{teo:canonicity}, we have the
following fact.

\begin{theorem}%
    \label{teo:canonicity-technique}
    Let \allalpha{},\allbeta{} denote arbitrary sets of \T-atoms with Boolean
    abstraction \allA{},\allB{} respectively.
    Consider some canonical form of \bdds{} on some canonicity condition
    \criterion{\allA}.
    Let \Tbdd{} with canonicity condition
    \criterion{\allalpha{}} be such that, for all sets \allalpha{},
    \allbeta{} and for all formulas \via{}:
    \begin{eqnarray}
        \label{eq:canonicity-technique}
        \Tbdd{}(\via{})\defas   \btot{}
        (\bdd(\viap{}\wedge \exists\allB.\hspace{-.7cm}\bigwedge_{\Cab{l}\in\ETLEMMAS{\vi}} \hspace{-.7cm}\Cabp{l})).
    \end{eqnarray}
    Then the \Tbdds{} are \T-canonical.
\end{theorem}

\section{Building Canonical \Tbdds{}}%
\label{sec:tdds}
\begin{figure}[t] 
  \begin{center} 
\scalebox{0.71}{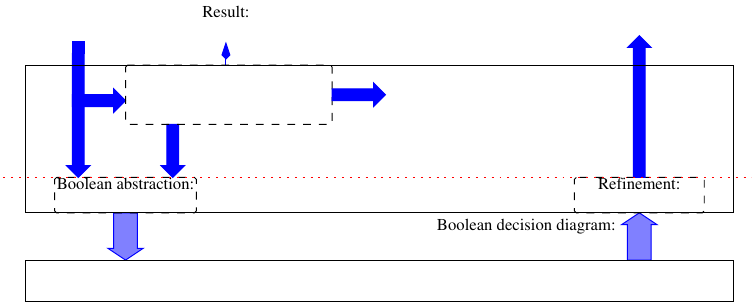}
  \end{center} 
  \vspace{-.5cm}
  \scalebox{0.8}{
    \begin{minipage}{0.4\textwidth}
      {\tt
        \begin{tabbing}
          \hspace{.3cm} \= \hspace{.3cm} \= \hspace{.3cm} \= \kill
          \\
          \Tbdd{}  {\sf T-DD\_Compiler}(\T-formula $\via{}$) \{ \\
          \> {\bf if}  (AllSMT\_Solver($\via{}$) == \unsatres)  \\
          \>\> {\bf then return} {\sf DD-False}; \\
          \> \green{// $\theorycl$ are the \T-lemmas stored by AllSMT\_Solver} \\
          \> $\mathcal{DD}^p$=DD\_Compiler$(\inputboolcl)$; \hspace{.8cm}\
          \\
          \> $\mathcal{DD}$=$\proptofol(\mathcal{DD}^p)$;\\
          \>{\bf return} $\mathcal{DD}$; \\
          \}\> 
        \end{tabbing}
      }
    \end{minipage}
  }
%
\vspace{.4cm}
\caption{
Schema of the \T-knowledge compiler: architecture (above)
  and algorithm (below). \\ \\
}\label{fig:compiler-algo} 
\end{figure}
Given some background theory \T and given some form of knowledge compiler for Boolean formulas
into some form of Boolean
decision diagrams 
(e.g., \obdds, \sdds, \ldots), \Cref{teo:canonicity-technique},
suggests us an
easy way to implement a compiler of an SMT formula into a \T-canonical
\Tbdd{}. 

\subsection{General Ideas}
The procedure is reported in \Cref{fig:compiler-algo}.
The input \T-formula \via{} is first fed to an AllSMT solver which enumerates the set 
$\CTTA{\via{}}\defas\set{\etaa{1},\ldots,\etaa{N}}$
of 
\T-satisfiable total assignments propositionally
satisfying \via{}.
To do that, the AllSMT solver has to produce a set
$\ETLEMMAS{\via{}}\defas\theoryclab{}$ 
of \T-lemmas that rule out all the \T-unsatisfiable assignments
in $\ITTA{\via{}}$.
SMT solvers like \mathsat{} can produce these \T-lemmas as output.

We ignore the $\eta_i$s and we conjoin the
\T-lemmas to $\via{}$.%
\footnote{In principle, we could feed the \bdd{} package directly the
  disjunction of all assignments $\eta$ in \CTTA{\via{}}. In practice, this
  would be extremely inefficient, since the $\eta$s are all total
  assignments and there is a large amount of them. Rather, the
  \T-lemmas typically  involve only a very small subset of \T-atoms,
  and thus each \T-lemma of length $k$ rules out up to $2^{|\allalpha|-k}$ \T-inconsistent
  total assignments in \ITTA{\via{}}. }
We apply the Boolean abstraction and feed 
\inputboolcl{} to a Boolean \bdd{}-Compiler, which returns one
decision diagram $\mathcal{DD}^p$, which is equivalent to
\inputboolcl{} in the Boolean space. 
$\mathcal{DD}^p$ is then mapped back into
a \Tbdd{} $\mathcal{DD}$ via \proptofol{}.  
Several DD-compilers support the transformation $\exists\allB.\vi$, which is referred to as \textit{forgetting} or \textit{projection}~\cite{darwicheKnowledgeCompilationMap2002}.





\begin{figure}[t]
\centering
  \begin{tikzpicture}[scale=0.8, transform shape, every node/.style={rectangle, draw, minimum size=6mm}, circle_node/.style={circle, draw, minimum size=6mm, rounded corners=4pt}]
    \node (A1) at (0,0) {$A_1$};
    \node (A2) at (1.5,-1) {$A_2$};
    \node (A3) at (0,-2) {$A_3$};
    \node[circle_node] (F) at (-1.5,-3) {$\bot$};
    \node[circle_node] (T) at (1.5,-3) {$\top$};
    \draw (A1) -- (A2);
    \draw[dashed] (A1) -- (F);
    \draw (A2) -- (T);
    \draw[dashed] (A2) -- (A3);
    \draw (A3) -- (T);
    \draw[dashed] (A3) -- (F);
\end{tikzpicture}
\hspace{1cm}
\begin{tikzpicture}[scale=0.8, transform shape, every node/.style={rectangle, draw, minimum size=6mm}, circle_node/.style={circle, draw, minimum size=6mm, rounded corners=4pt}]
    \node (A1) at (0,0) {$x \leq 0$};
    \node (A2) at (1.5,-1) {$x \geq 1$};
    \node (A3) at (0,-2) {$x \leq 2$};
    \node[circle_node] (F) at (-1.5,-3) {$\bot$};
    \node[circle_node] (T) at (1.5,-3) {$\top$};
    \draw (A1) -- (A2);
    \draw[dashed] (A1) -- (F);
    \draw (A2) -- (T);
    \draw[dashed] (A2) -- (A3);
    \draw (A3) -- (T);
    \draw[dashed] (A3) -- (F);
    \draw[red, line width=1.5pt] ($(A1)!0.3!(A2)$) -- ($(A1)!0.7!(A2)$);
    \draw[red, line width=1.5pt] ($(A2)!0.15!(T)$) -- ($(A2)!0.85!(T)$);
\end{tikzpicture}
\\
\begin{tikzpicture}[scale=0.8, transform shape, every node/.style={rectangle, draw, minimum size=6mm}, circle_node/.style={circle, draw, minimum size=6mm, rounded corners=4pt}]
    \node (A1) at (0,0) {$A_1$};
    \node (A2) at (0.75,-1) {$A_2$};
    \node (A3) at (1.5,-2) {$A_3$};
    \node[circle_node] (F) at (-1.5,-3) {$\bot$};
    \node[circle_node] (T) at (1.5,-3) {$\top$};
    \draw (A1) -- (A2);
    \draw[dashed] (A1) -- (F);
    \draw[dashed] (A2) -- (A3);
    \draw (A2) -- (F);
    \draw (A3) -- (T);
    \draw[dashed] (A3) -- (F);
\end{tikzpicture}
\hspace{1cm}
\begin{tikzpicture}[scale=0.8, transform shape, every node/.style={rectangle, draw, minimum size=6mm}, circle_node/.style={circle, draw, minimum size=6mm, rounded corners=4pt}]
    \node (A1) at (0,0) {$x \leq 0$};
    \node (A2) at (0.75,-1) {$x \geq 1$};
    \node (A3) at (1.5,-2) {$x \leq 2$};
    \node[circle_node] (F) at (-1.5,-3) {$\bot$};
    \node[circle_node] (T) at (1.5,-3) {$\top$};
    \draw (A1) -- (A2);
    \draw[dashed] (A1) -- (F);
    \draw[dashed] (A2) -- (A3);
    \draw (A2) -- (F);
    \draw (A3) -- (T);
    \draw[dashed] (A3) -- (F);
\end{tikzpicture}
\vspace{.4cm}
  \caption{
    Top left: \obdd{} of $\vip\defas \Aone \wedge (\Atwo\vee \Athree)$;
    Top right: refinement of the \obdd{} of $\vi\defas \cone \wedge (\ctwo\vee \cthree)$; 
    Bottom left: \obdd{} of $\vip\wedge C_1^p\bequiv (\Aone\wedge \neg
    \Atwo \wedge \Athree)$; 
    Bottom right: refinement of the \obdd{} of $\vip\wedge
    C_1^p\Tequiv\cone\wedge \neg \ctwo \wedge \cthree$.
\\ \\}\label{fig:obdd}
\end{figure}

\begin{example}
  Let $\allalpha\defas \set{\cone,\ctwo,\cthree}$.
Consider the \larat-formula
$\vi\defas \cone \wedge (\ctwo\vee \cthree)$,
and its Boolean abstraction
$\vip\defas \Aone \wedge (\Atwo\vee \Athree)$.
Assume the order \set{\cone,\ctwo,\cthree}.
The \obdd{} of \vip{} and its refinement are
reported in \Cref{fig:obdd}, top left and right.
Notice that the latter has one \T-inconsistent branch
\set{\cone,\ctwo} (in red). \\
The AllSMT solver can enumerate the satisfying assignments:\\
$\begin{array}{lll}
\rho_1  &\defas & \cone\wedge\pos\ctwo\wedge\pos\cthree\\
\rho_2  &\defas & \cone\wedge\pos\ctwo\wedge\neg\cthree\\
\eta_1  &\defas & \cone\wedge\neg\ctwo\wedge\pos\cthree\\
\end{array}$\\
causing the generation of the following \T-lemma to rule out $\rho_1,\rho_2$:\\
$C_1\defas\neg\cone\vee\neg\ctwo$, whose Boolean abstraction is
$C_1^p\defas\neg\Aone\vee\neg\Atwo$. (Since $\allbeta=\emptyset$
here, there is no need to existentially quantify \allB.)

Passing $\vip\wedge C_1^p\bequiv(\Aone\wedge\neg\Atwo\wedge\Athree)$ to an
\obdd compiler, the \obdd returned is the one in
\Cref{fig:obdd}, bottom left, corresponding to the \Tobdd on
bottom right. Notice that the \T-inconsistent branch has been
removed, and that there is no
    \T-inconsistent branch left.
\end{example}



  
  \begin{remark}
    We stress the fact that our approach is {\em not} a form of eager SMT
encoding for \bdd construction. The latter consists in enumerating a
priori all possible \T-lemmas which can be constructed on top of the
\T-atom set \allalpha, regardless of the formula \via{}~\cite{barrettSatisfiabilityModuloTheories2021}. Except for very simple theories like \eq{} or \euf{}, this causes a
huge amount of \T-lemmas. 
With our approach, which is inspired instead
by the
``lemma-lifting'' approach for SMT unsat-core extraction~\cite{cimattiComputingSmallUnsatisfiable2011} and MaxSMT~\cite{cimattiModularApproachMaxSAT2013},
%
only the  \T-lemmas which
are needed to rule out the \T-inconsistent truth assignments in
\ITTA{\vi}, are generated on demand by the AllSMT solver.    
\end{remark}

\subsection{About Canonicity}

\begin{figure}[t]
   \begin{tikzpicture}[scale=0.8, transform shape, every node/.style={rectangle, draw, minimum size=6mm}, circle_node/.style={circle, draw, minimum size=6mm, rounded corners=4pt}]
    \node (A0) at (0,0) {$A_0$};
    
    \node (A1-1) at (-1.2,-1) {$A_1$};
    \node (A1-2) at (1.2,-1) {$A_1$};
    
    \node (A2-1) at (-2,-2) {$A_2$};
    \node (A2-2) at (0.5,-2) {$A_2$};
    \node (A2-3) at (2,-2) {$A_2$};
    
    \node[circle_node] (F) at (-1,-4) {$\bot$};
    \node[circle_node] (T) at (1,-4) {$\top$};
    \draw (A0) -- (A1-2);
    \draw[dashed] (A0) -- (A1-1);
    \draw[dashed] (A1-1) -- (A2-1);
    \draw (A1-1) -- (T);
    \draw[dashed] (A1-2) -- (A2-2);
    \draw (A1-2) -- (A2-3);
    \draw[dashed] (A2-1) -- (F);
    \draw (A2-1) -- (T);
    \draw[dashed] (A2-2) -- (F);
    \draw (A2-2) -- (T);
    \draw[dashed] (A2-3) -- (T);
    \draw (A2-3) -- (F);
\end{tikzpicture}
\hspace{0.5cm}
  \begin{tikzpicture}[scale=0.8, transform shape, every node/.style={rectangle, draw, minimum size=6mm}, circle_node/.style={circle, draw, minimum size=6mm, rounded corners=4pt}]
    \node (A0) at (0,0) {$y \leq 0$};
    
    \node (A1-1) at (-1.2,-1) {$x \leq 0$};
    \node (A1-2) at (1.2,-1) {$x \leq 0$};
    
    \node (A2-1) at (-2,-2) {$x = 1$};
    \node (A2-2) at (0.5,-2) {$x = 1$};
    \node (A2-3) at (2,-2) {$x = 1$};
    
    \node[circle_node] (F) at (-1,-4) {$\bot$};
    \node[circle_node] (T) at (1,-4) {$\top$};
    \draw (A0) -- (A1-2);
    \draw[dashed] (A0) -- (A1-1);
    \draw[dashed] (A1-1) -- (A2-1);
    \draw (A1-1) -- (T);
    \draw[dashed] (A1-2) -- (A2-2);
    \draw (A1-2) -- (A2-3);
    \draw[dashed] (A2-1) -- (F);
    \draw (A2-1) -- (T);
    \draw[dashed] (A2-2) -- (F);
    \draw (A2-2) -- (T);
    \draw[dashed] (A2-3) -- (T);
    \draw (A2-3) -- (F);
\end{tikzpicture}
\vspace{.4cm}
  \caption{
    \obdds for $(\neg \Azero \vee (\Aone \vee \Atwo)) \wedge (\Azero
    \vee (\Aone \oplus \Atwo))$ and \newline
    \Tobdd for $(\neg \azero \vee (\aone \vee \atwo)) \wedge (\azero \vee (\aone \oplus \atwo))$. \\ \\
    }\label{fig:canonicity}
\end{figure}

\begin{figure}[t]
  \centering
  \begin{tikzpicture}[scale=0.8, transform shape, every node/.style={rectangle, draw, minimum size=6mm}, circle_node/.style={circle, draw, minimum size=6mm, rounded corners=4pt}]
    \node (A) at (0,0) {$x\leq y$};
    \node (C1) at (-1,-2) {$y \leq z$};
    \node (C2) at (1,-2) {$y \leq z$};
    \node[circle_node] (T) at (-1,-3) {$\top$};
    \node[circle_node] (F) at (1,-3) {$\bot$};
    \draw (A) -- (C1);
    \draw[dashed] (A) -- (C2);
    \draw (C1) -- (T);
    \draw[dashed] (C1) -- (F);
    \draw (C2) -- (F);
    \draw[dashed] (C2) -- (T);
    
\end{tikzpicture}
\hspace{1cm}
\begin{tikzpicture}[scale=0.8, transform shape, every node/.style={rectangle, draw, minimum size=6mm}, circle_node/.style={circle, draw, minimum size=6mm, rounded corners=4pt}]
    \node (A) at (0,0) {$x\leq y$};
    \node (B) at (-1,-1) {$x \le z$};
    \node (C1) at (-1.5,-2) {$y \leq z$};
    \node (C2) at (1.5,-2) {$y \leq z$};
    \node[circle_node] (T) at (-1,-3) {$\top$};
    \node[circle_node] (F) at (1,-3) {$\bot$};
    \draw (A) -- (B);
    \draw[dashed] (A) -- (C2);
    \draw[dashed] (B) -- (F);
    \draw (B) -- (C1);
    \draw (C1) -- (T);
    \draw[dashed] (C1) -- (F);
    \draw (C2) -- (F);
    \draw[dashed] (C2) -- (T);
    
\end{tikzpicture}
\vspace{.4cm}
\caption{
  Let $\criterion{\allalpha}\defas\set{(x \leq y),(x
      \leq z),(y \leq z)}$. The DDDs/LDDs for the \T-formulas $\phi_1 = (x \leq y) \iff (y
    \leq z)$ and $\phi_2 = \phi_1 \wedge (\neg (x \leq y) \vee (x \leq
    z) \vee \neg (y \leq z))$ are reported above. (On these formulas, the output of DDD and
    LDD is the same).
  Notice that $(\neg (x \leq y) \vee (x \leq
    z) \vee \neg (y \leq z))$ is \T-valid so that $\phi_1 \Tequiv \phi_2$, but the diagrams are different. \\
  \\} 
  \label{fig: DDD-LDD-not-canonical}
\end{figure}

%
A big advantage of our approach is that
\Cref{teo:canonicity} guarantees that,  given an ordered set \allalpha{} of
atoms, {\em we produce
  canonical \Tobdds}.
The importance of \T-canonicity is shown in the following example.
\begin{example}
  \label{ex:canonical}
  Consider $\criterion{\allalpha}\defas\set{\azero,\aone,\atwo}$ and let
  $\vi\defas (\neg \azero \vee \vi_{1}) \wedge (\azero \vee \vi_{2})$,
  which contains $\vi_{1}\defas(\aone \vee \atwo)$ and $\vi_{2}\defas(\aone
  \oplus \atwo)$ of \Cref{ex:proposition1},  which are
  \T-equivalent, but which may not be recognized as
  such by previous forms of \Tobdds. If this is the case, the final
  \Tobddof{\vi} contains
  \Tobddof{\vi_{1}} and \Tobddof{\vi_{2}}, without realizing they are
  \T-equivalent.
  For example, if \Tobddof{\vi_{1}} and \Tobddof{\vi_{2}} are these in
  \Cref{fig:noncanonical} top right and bottom right
  respectively, then \Tobddof{\vi}
  is represented in \Cref{fig:canonicity} right.

  With our approach, instead, since $\vi\Tequiv\vi_{1}\Tequiv\vi_{2}$
  and thanks to \Cref{teo:canonicity}, 
  we produce the
  \obdd and \Tobdd of \Cref{fig:noncanonical}, bottom left and
  right.
  In fact, $\vi$ is \T-equivalent to  $\vi_{1}$ and to
  $\vi_{2}$, because it is in the form
  $(\neg\beta_1\vee\vi_{1})\wedge(\beta_1\vee\vi_{2})$ where $\vi_{1}\Tequiv\vi_{2}$.
\end{example}


\begin{remark}  
The notion of \T-canonicity, like that of Boolean
canonicity, assumes that the formulas \via{} and
\viaprime{}  {\em are compared on the same (super)set of \T-atoms
  \allalpha{}}.
Therefore, in order to compare two formulas on different atom sets,
$\viab{}$ and $\viabprime{}$, we need to consider them as
formulas on the union of atoms sets: $\viabb{}$ and $\viabbprime{}$.
If so, our technique produces also the necessary \T-lemmas which 
relate $\allalpha,\allbeta,\allbetaprime$.
\end{remark}

\begin{example}
  In order to compare 
$\psi_1\defas(x=0)\wedge(y=1)$ and
$\psi_2\defas(x=0)\wedge(y=x+1)$, we need to consider them on the
\T-atoms \set{(x=0),(y=1),(y=x+1)}. \\
If so, with our procedure the AllSMT solver produces
the
\T-lemmas
$C_1\defas\neg(x=0)\vee\neg(y=1)\vee(y=x+1)$ for $\psi_1$ and
$C_2\defas\neg(x=0)\vee\neg(y=x+1)\vee(y=1)$ for $\psi_2$, and builds the \bdds of
$\psi_1^p\wedge C_1^p$ and
$\psi_2^p\wedge C_2^p$ which are equivalent, so that the two \Tbdds
are identical, and are identical to that of $(x=0)\wedge(y=1)\wedge(y=x+1)$.
\end{example}

Ensuring \T-canonicity is not straightforward. For example, 
DDDs~\cite{mollerDifferenceDecisionDiagrams1999} and LDDs~\cite{chakiDecisionDiagramsLinear2009} are not \T-canonical: e.g., in
the example in \Cref{fig:
  DDD-LDD-not-canonical} both produce different \Tbdds{} for two
\T-equivalent formulas. Additionally, LDDs are not even
\T-semicanonical: in \Cref{fig: LDD-not-semi} we show an example
where LDDs' theory-specific simplifications fail to reduce a
\T-inconsistent formula to the node $\bot$.

\begin{figure}[t]
  \centering
  \begin{tikzpicture}[scale=0.8, transform shape, every node/.style={rectangle, draw, minimum size=6mm}, circle_node/.style={circle, draw, minimum size=6mm, rounded corners=4pt}]
    \node (XZ) at (0,0) {$x - z \leq -3$};
    \node (YX) at (1,-1) {$y - x \leq 2$};
    \node (ZY) at (2,-2) {$z - y \leq -1$};
    \node[circle_node] (F) at (-1,-3) {$\bot$};
    \node[circle_node] (T) at (3,-3) {$\top$};
    \draw (XZ) -- (YX);
    \draw[dashed] (XZ) -- (F);
    \draw (YX) -- (ZY);
    \draw[dashed] (YX) -- (F);
    \draw (ZY) -- (T);
    \draw[dashed] (ZY) -- (F);
\end{tikzpicture}
\vspace{.4cm}
  \caption{LDD for $\phi_3 \defas (x - z \leq -3) \vee (y - x \leq 2)
    \vee (z - y \leq -1)$. Notice that $\phi_3$ is \T-unsatisfiable,
    yet the LDD is not the $\bot$ node. \\ \\}
  \label{fig: LDD-not-semi}
\end{figure}

\section{A Preliminary Empirical Evaluation}%
\label{sec:expeval}

To test the feasibility of our approach, we developed a prototype of
the \Tbdd{} generator described in  the algorithm in
\sref{sec:tdds}. Our tool, coded in Python using
PySMT~\cite{garioPySMTSolveragnosticLibrary2015} for parsing and
manipulating formulas, leverages: (i) CUDD~\cite{somenzi2009cudd} for
\obdd{} generation; (ii) SDD~\cite{darwicheSDDNewCanonical2011} for
SDD generation; (iii) \mathsatfive{}~\cite{mathsat5_tacas13} for
AllSMT enumeration and theory lemma generation. 

The benchmarks and the results are available at~\cite{michelutti_2024_benchmark}, and the source code of the tool at~\cite{michelutti_2024_tool}. See also \url{https://github.com/MaxMicheluttiUnitn/TheoryConsistentDecisionDiagrams} and \url{https://github.com/MaxMicheluttiUnitn/DecisionDiagrams} for an up-to-date version of the tool.

\subsection{Comparison With Other Tools}

As reported in \sref{sec:intro}, the existing toolsets in the field
are very limited. Indeed, for \Tobdds{} only LDD~\cite{chakiDecisionDiagramsLinear2009} 
have a public and directly usable implementation. (See the analysis of tools in~\Cref{sec:apdx:rw}.) 
Moreover, LDD's implementation is confined to \tvpi{} over real or integer variables.
For \Tsdds, the implementation of XSDD is intricately tailored for Weighted Model Integration problems, making the extraction of \Tsdd{} from its code non-trivial.

Thus, we conducted a comparative analysis of our tool against the following tools:
\begin{enumerate*}[label=(\roman*)]
\item \textit{Abstract \obdd}, baseline \Tobdd{} obtained from the refinement of the \obdd{} of the Boolean abstraction built with CUDD;\@ 
\item \textit{Abstract \sdd}, baseline \Tsdd{} obtained from the refinement of the \sdd{} of the Boolean abstraction built with SDD.\@ We remark that, as indicated in~\cite{dosmartiresExactApproximateWeighted2019}, the XSDD construction aligns with \textit{Abstract \sdd};\@
\item \textit{LDD} from~\cite{chakiDecisionDiagramsLinear2009}.
\end{enumerate*}
%
Our analysis focuses on two metrics:
\begin{enumerate*}[label=(\roman*)]
\item the time required to compile the formulas, and
\item the number of nodes in the diagrams.
\end{enumerate*}
%
%
We assume uniformity in variable
ordering (for \obdds) and v-tree (for \sdds) across the tools. 
\begin{figure}[t]
    \centering
 \begin{subfigure}[t]{0.48\columnwidth}
     \centering
     \includegraphics[width=\textwidth]{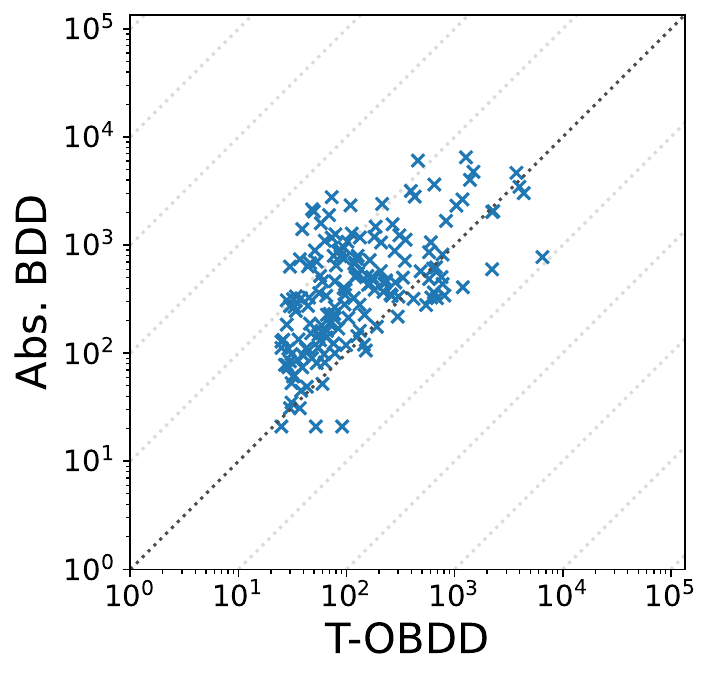}%
     \label{fig:plot sintetici LRA BDD vs TBDD size}
 \end{subfigure}
 \begin{subfigure}[t]{0.48\columnwidth}
     \centering
     \includegraphics[width=\textwidth]{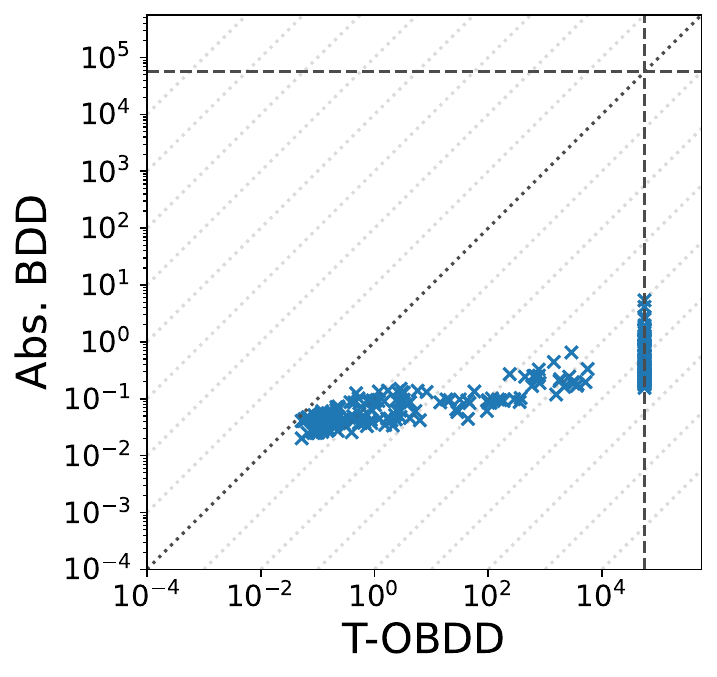}%
     \label{fig:plot sintetici LRA BDD vs TBDD time}
 \end{subfigure} \\
 \begin{subfigure}[t]{0.48\columnwidth}
     \centering
     \includegraphics[width=\textwidth]{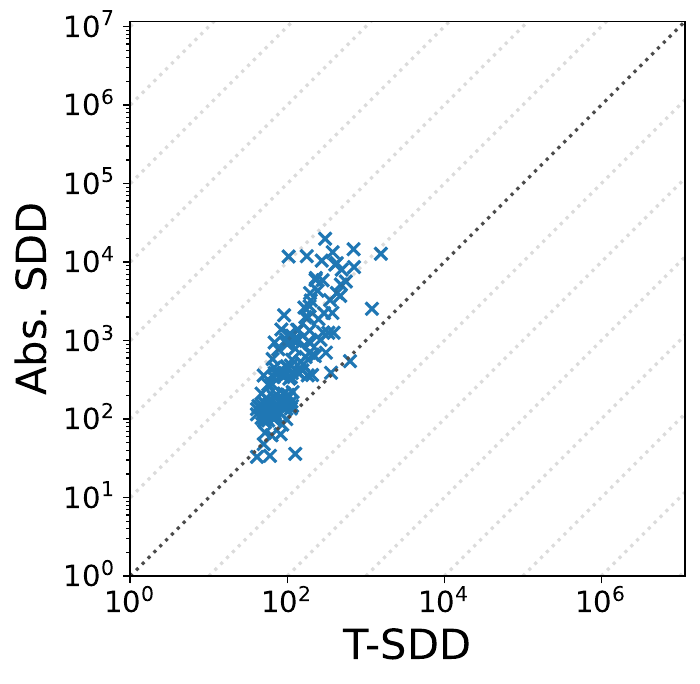}
     \caption{Number of nodes}%
     \label{fig:plot sintetici LRA SDD vs TSDD size}
 \end{subfigure}
 \begin{subfigure}[t]{0.48\columnwidth}
     \centering
     \includegraphics[width=\textwidth]{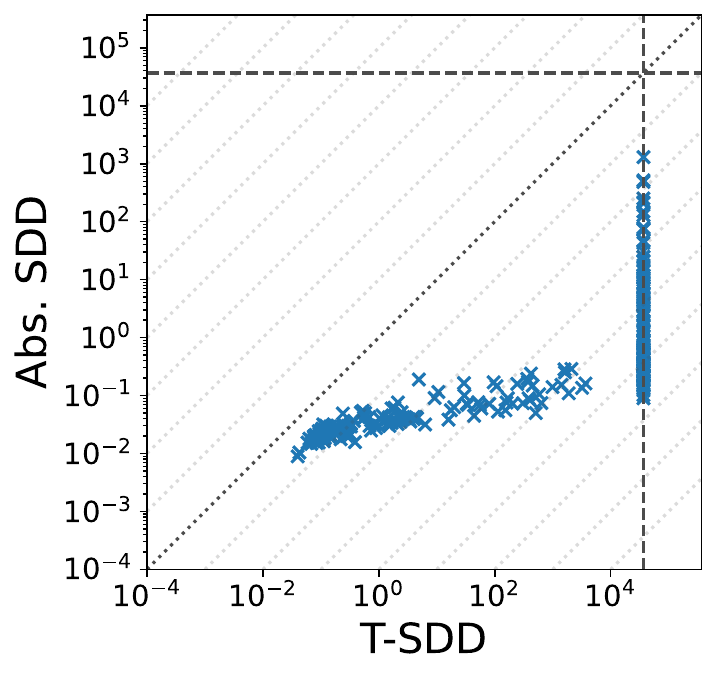}
    \caption{Computational time}%
     \label{fig:plot sintetici LRA SDD vs TSDD time}
 \end{subfigure}
 \vspace{.4cm}
 \caption{Results obtained on synthetic \larat{} benchmarks (250 problems), comparing number of nodes (left) and computational times (right). Timeouts on the horizontal and vertical lines. \Tobdd{} timeouts: 79. \Tsdd{} timeouts: 111. \\ \\}\label{fig:sinteticiLRA}
\end{figure}

We wish to stress the fact that the procedures against which we empirically compare cannot be fairly considered as ``competitors'' since none of them is \T-canonical, nor
even \T-semicanonical. Since canonicity is an intrinsic source of hardness, we expect the time computation of our tool to be higher than the competitors. Canonicity, however, provides several properties that the other techniques are not guaranteed to satisfy, and we will report them in our results.
\subsection{Benchmark}

Due to the limited literature on \Tbdds, 
there is a scarcity of benchmarks available. As a first
step, we tested our tool on a subset of SMT-LIB benchmark
problems. The main issue is that SMT-LIB problems are thought for SMT
solving, and not for knowledge compilation. As a result, most of the
problems are UNSAT or too difficult to compile into a \Tbdd{} in a feasible
amount of time by any tool. Hence, we generated problems inspired by the Weighted
Model Integration application, drawing inspiration from~\cite{dosmartiresExactApproximateWeighted2019,kolbHowExploitStructure2020}. In this context, we consulted
recent papers on the topic~\cite{spallittaEnhancingSMTbasedWeighted2024a} and crafted a
set of synthetic benchmarks accordingly. We set the weight function to
1 to prioritize the generation of the \Tbdd{} of the support
formula, and adjusted the generation code to align with
theories supported by the competitors (i.e., \tvpi{} for LDD
 and
\larat{} for XSDD).

\begin{figure}[t!]
    \centering
 \begin{subfigure}[b]{0.48\columnwidth}
     \centering
     \includegraphics[width=\textwidth]{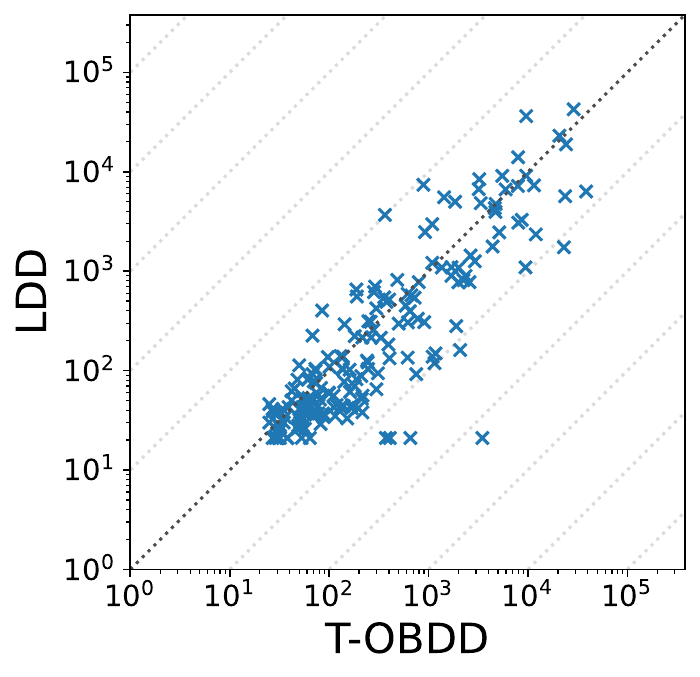}%
     \label{fig:plot sintetici LDD compatibili LDD vs TBDD size}
 \end{subfigure}
 \begin{subfigure}[b]{0.48\columnwidth}
     \centering
     \includegraphics[width=\textwidth]{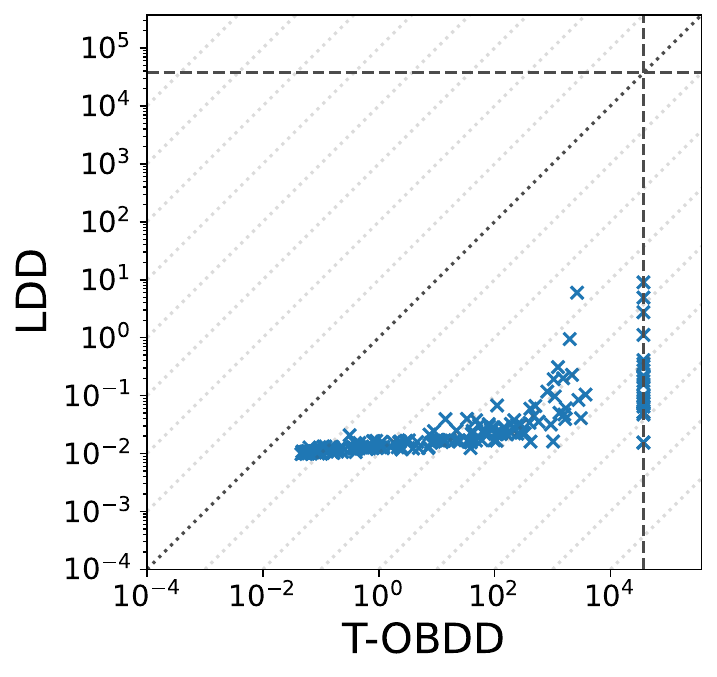}%
     \label{fig:plot sintetici LDD compatibili LDD vs TBDD time}
 \end{subfigure} \\
  \begin{subfigure}[b]{0.48\columnwidth}
     \centering
     \includegraphics[width=\textwidth]{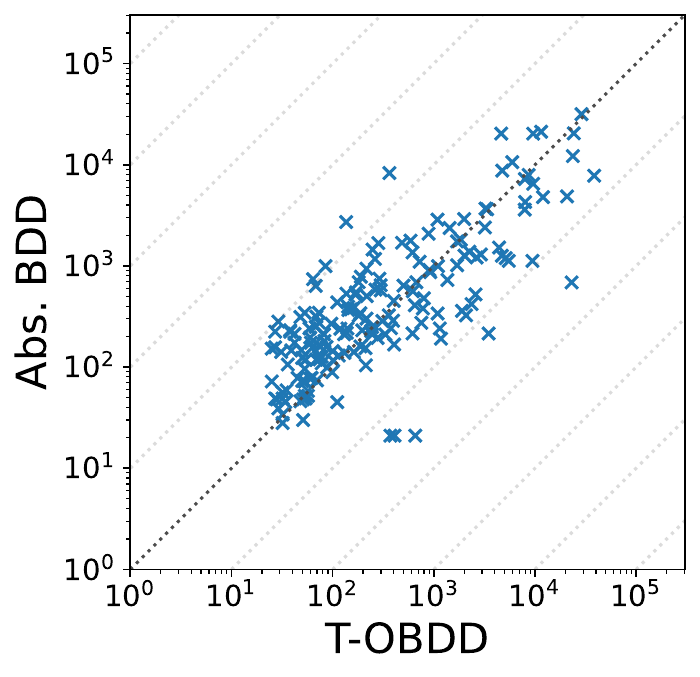}%
     \label{fig:plot sintetici LDD compatibili BDD vs TBDD size}
 \end{subfigure}
 \begin{subfigure}[b]{0.48\columnwidth}
     \centering
     \includegraphics[width=\textwidth]{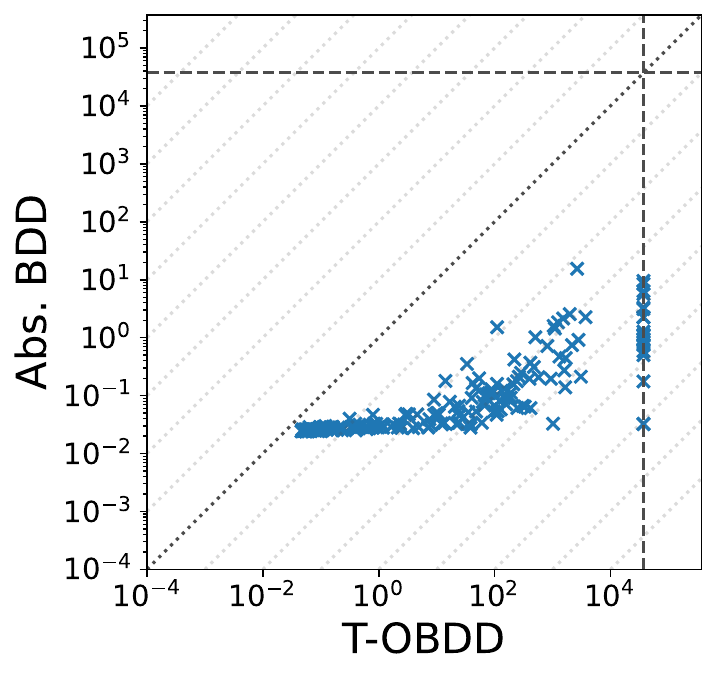}%
     \label{fig:plot sintetici LDD compatibili BDD vs TBDD time}
 \end{subfigure} \\
  \begin{subfigure}[b]{0.48\columnwidth}
     \centering
     \includegraphics[width=\textwidth]{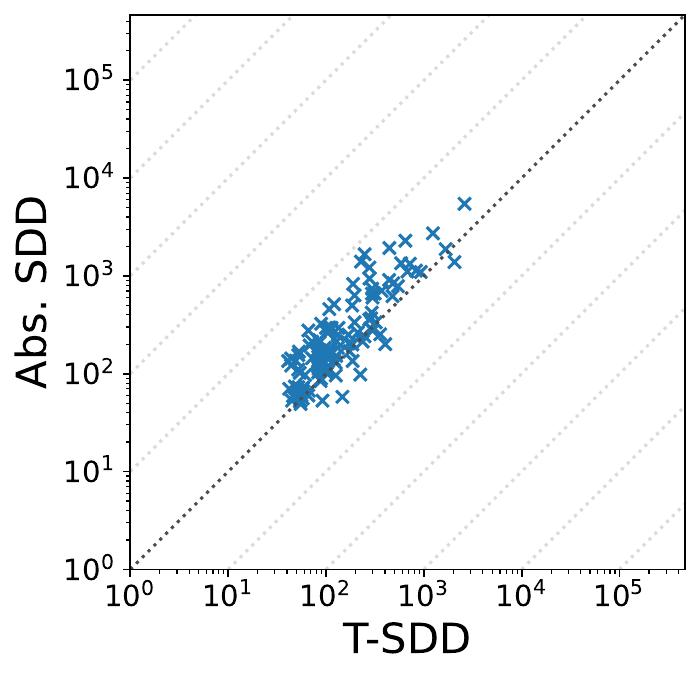}
    \caption{Number of nodes}%
     \label{fig:plot sintetici LDD compatibili SDD vs TSDD size}
 \end{subfigure} 
 \begin{subfigure}[b]{0.48\columnwidth}
     \centering
     \includegraphics[width=\textwidth]{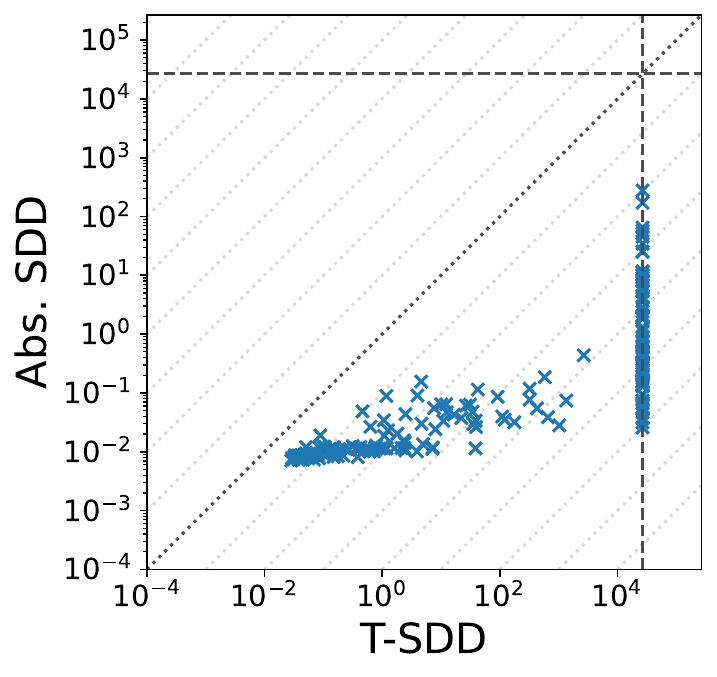}
    \caption{Computational time}%
     \label{fig:plot sintetici LDD compatibili SDD vs TSDD time}
 \end{subfigure}
 \vspace{.4cm}
 \caption{Results obtained on synthetic \tvpi{} benchmarks (200 problems), comparing number of nodes (left) and computational times (right). Timeouts on the horizontal and vertical lines. LDD timeouts: 0. \Tobdd{} timeouts: 22. \Tsdd{} timeouts: 81.\\ \\}%
 \label{fig:sinteticiTVPI}
\end{figure}

\subsection{Results}

Figures~\ref{fig:sinteticiLRA} and~\ref{fig:sinteticiTVPI} show the comparison of our algorithm ($x$-axis) against all baseline solvers ($y$-axis). The results are shown through scatter plots, comparing the size of the generated \Tbdds{} and the taken computational time. We set the timeout to 3600s for AllSMT computation, and additional 3600s for \Tbdd{} generation.
Notice that both axes are log-scaled. On the one hand, the plots show that
our algorithms have longer computational times compared to the other tools. This outcome is not surprising, given the
additional overhead associated with enumerating the
lemmas via AllSMT and performing Boolean existential quantification. 
On the other hand, our tools generate smaller \Tbdds{}, which is particularly noticeable for \Tsdds{}. 

Our tools offer several distinctive advantages that set them apart within the field. Notably, these advantages may not be readily discernible from scatter plots or other visualization methods. 

   The \Tbdds{} built with our approach ensure that every extension of a partial assignment leading to the $\top$ node represents a \T-consistent total assignment. Consequently, our algorithm stands as the sole contender capable of performing \#SMT, aligning with the definition of \#SMT proposed in~\cite{phanModelCountingModulo2015}. This characteristic holds substantial implications for various applications, particularly in fields like Quantitative Information Flow, where precise enumeration is crucial.

    Furthermore, benchmarks from the SMT-LIB, predominantly comprising UNSAT instances, proved our capability to identify \T-inconsistent formulas and condense them into a single $\bot$ node. In contrast, LDD do not generate a $\bot$ \Tbdd{} for these formulas, highlighting once again their lack in achieving \T-semicanonicity.
  
    Finally, our algorithm supports the combination of theories and addresses theories not supported by other available implementations. In~\cite{michelutti_2024_tool}, we provide a collection of problems spanning various theories, all of which are compatible with our implementation. Notably, our tool is the only public tool capable of generating theory decision diagrams for these problem domains.

\section{Conclusions and Future Work}%
\label{sec:concl}

In this paper, we have investigated the problem of leveraging Boolean
decision diagrams (\bdds{}) to SMT level (\Tbdds{}). We have presented a
general theory-agnostic and \bdd{}-agnostic formal framework for \Tbdds{}.
We have shown a straightforward way to leverage \bdds{} to \Tbdds{} by simply combining an AllSMT solver and a \bdd{} package, both used as black boxes.
This approach requires little effort to implement, since it does not
require to modify the code of the AllSMT solver and of the \bdd{}
package,  and is very general,
since it can be applied to any theory supported by the AllSMT solver
and combinations thereof, 
and to any \bdd{} with a compiler admitting Boolean
existential quantification.
Importantly, this technique has a fundamental feature: it allows leveraging canonical \bdds{} into \T-canonical \Tbdds{}.
To the best of our knowledge, this is the first case 
 of provably canonical \Tbdds{} in the literature.
We have implemented our approach on top of the \mathsatfive{} AllSMT solver
and of both \obdd{} and \sdd{} packages, and shown empirically its
effectiveness.

  
This work opens several research directions.

From a theoretical viewpoint, we are going to investigate how
\Tbdds{} can be effectively composed 
and how querying can be performed;
also, we plan to extend our analysis to other forms of \bdds{}, and on
NNF formulas in general, investigating how their properties can be preserved
by leveraging to SMT level. 
Of particular interest are d-DNNFs, whose extension to SMT level is a recent topic of research~\cite{derkinderenTopDownKnowledgeCompilation2023}. 

From a practical viewpoint, our approach currently suffers from two main
bottlenecks:
\begin{enumerate*}[label=(\alph*)]
    \item the need to perform AllSMT upfront, and
    \item the need to perform Boolean existential quantification to remove the extra \T-atoms.
\end{enumerate*}
For the former, we plan to investigate alternative and
less-expensive ways to enumerate \T-lemmas ruling out \T-inconsistent
assignments, such as exploiting AllSAT strategy without introducing blocking clauses~\cite{spallittaDisjointPartialEnumeration2024}. For the latter, we plan to investigate alternative SMT
techniques which reduce or even eliminate the presence of novel
\T-atoms in the \T-lemmas. 

From an application viewpoint, we plan to use our \Tsdds{} package
for Weighted Model Integration (WMI), with the idea of merging the
best features of AllSMT-based WMI~\cite{morettin-wmi-ijcar17,morettin-wmi-aij19,spallittaSMTbasedWeightedModel2022,spallittaEnhancingSMTbasedWeighted2024a}, and
those of KC-based WMI~\cite{dosmartiresExactApproximateWeighted2019,kolbHowExploitStructure2020}. Since in WMI the expensive part relies on the computation of integrals, the fewer truth assignments are represented, the fewer integrals must be computed (and the faster the computation). Our \Tsdds{} prune in advance \T-inconsistent assignments, which would give no contribution to the integral or number of solutions. Moreover, canonicity typically allows for obtaining more compact diagrams, which in turn can speed up the computation of integrals and the counting of solutions.
Also, theory OBDDs have been proposed as an alternative to SMT-based techniques for checking verification properties on timed systems and software systems (e.g.,~\cite{mollerDifferenceDecisionDiagrams1999,cavadaComputingPredicateAbstractions2007, chakiDecisionDiagramsLinear2009,cimattiTighterIntegrationBDDs2010}). 




\begin{ack}
We acknowledge the support of the MUR PNRR project FAIR -- Future AI
Research (PE00000013), under the NRRP MUR program funded by the
NextGenerationEU.\@ The work was partially supported by the project
``AI@TN'' funded by the Autonomous Province of Trento.
The work was also funded in part by the European Union. Views and opinions
expressed are however those of the author(s) only and do not
necessarily reflect those of the European Union or the European Health
and Digital Executive Agency (HaDEA). Neither the European Union nor
the granting authority can be held responsible for them. Grant
Agreement no.\ 101120763 -- TANGO. \\
We thank Alberto Griggio for assistance with the MathSAT usage.
\end{ack}



\FloatBarrier
\bibliography{ecai-2024}

\newpage
\newpage
\appendix
\section{Appendix: Proofs of the theorems}
\label{sec:apdx:proofs}

\paragraph{Proof of \Cref{teo:basicprop}}
\begin{proof}
    Let
    $\Psi[\allalpha]\defas\Tbddof{\via{}}$ and
    $\Psi'[\allalpha]\defas\Tbddof{\viaprime{}}$. Then
    $\Psi[\allalpha]=\Psi'[\allalpha]$ if and only if
    $\Psi^p[\allalpha]=\Psi'^p[\allalpha]$. By Definition~\ref{def:tbdd},
    $\Psi^p[\allalpha]$ and $\Psi'^p[\allalpha]$ are \bdds{}, which are
    canonical by hypothesis.
    Thus $\Psi^p[\allalpha]=\Psi'^p[\allalpha]$ if and only if
    $\Psi^p[\allalpha]\equiv\Psi'^p[\allalpha]$, that is,
    if and only if $\Psi[\allalpha]\bequiv\Psi'[\allalpha]$.
\end{proof}

\paragraph{Proof of \Cref{teo:canonicity-sufficient}}
\begin{proof}
    Consider two \smtt{} formulas $\via{}$ and $\viaprime{}$.
    \\
    By \Cref{prop:assignmentsets}(d),
    $\via{}\Tequiv\viaprime{}$ if and only if
    $\CTTA{\vi{}}=\CTTA{\viprime{}}$.
    \\
    By the definition of  \CTTA{...} all the $\eta{}$s in \CTTA{...} are
    total on \allalpha{} and pairwise disjoint, so that $\CTTA{\vi{}}=\CTTA{\viprime{}}$ if
    and only if $\bigvee_{\eta_i\in\CTTA{\vi{}}}\eta_i\bequiv
        \bigvee_{\eta'_i\in\CTTA{\viprime{}}}\eta'_i$.
    \\
    Since
    $\Tbddof{\via{}}\bequiv \bigvee_{\eta_i\in\CTTA{\vi{}}}\eta_i$ and
    $\Tbddof{\viaprime{}}\bequiv\bigvee_{\eta'_i\in\CTTA{\viprime{}}}\eta'_i$,
    then
    we have that $\CTTA{\vi{}}=\CTTA{\viprime{}}$ if and only if $\Tbddof{\via{}}\bequiv
        \Tbddof{\viaprime{}}$.\\
    By \Cref{teo:basicprop}, $\Tbddof{\via{}}\bequiv
        \Tbddof{\viaprime{}}$ if and only if
    $\Tbddof{\via{}}=\Tbddof{\viaprime{}}$.
\end{proof}

\paragraph{Proof of \Cref{teo:teo2}}
\begin{proof}    
\Cref{teo:teo2} is a corollary of \Cref{teo:teo-ext}
(which we prove below) by setting $\allbeta\defas\emptyset$.
\end{proof}

\paragraph{Proof of \Cref{teo:canonicity-technique-noextraatoms}}
\begin{proof}    
\Cref{teo:canonicity-technique-noextraatoms} is a corollary of \Cref{teo:canonicity-technique}
(which we prove below) by setting $\allbeta\defas\emptyset$.
\end{proof}

\paragraph{Proof of \Cref{teo:teo-ext}}
\begin{proof}
    Since \ETLEMMAS{\vi} rules out \ITTA{\vi}, we have:
    \begin{eqnarray}
        \label{eq:teo3-hp3}
        &&\bigvee\limits_{\rhoa{j}\in\ITTA{\vi}}\hspace{-.6cm}\rhoa{j}\ \wedge\
        \bigwedge_{\Cab{l}\in\ETLEMMAS{\vi}} \hspace{-.6cm}\Cab{l}\
        \bequiv\ \bot,
        \\
        \mbox{i.e.:}    &&\bigvee\limits_{\rhoa{j}\in\ITTA{\vi}}\hspace{-.6cm}\rhoap{j}\ \wedge\
        \hspace{-.6cm}\bigwedge\limits_{\Cab{l}\in\ETLEMMAS{\vi}} \hspace{-.6cm}\Cabp{l}\
        \equiv\ \bot,
        \\
        \label{eq:teo3-hp3bool}
        \mbox{equiv.:}    &&\bigvee\limits_{\rhoa{j}\in\ITTA{\vi}}\hspace{-.6cm}\rhoap{j}\ \wedge\
        \exists \allB. \hspace{-.6cm}\bigwedge\limits_{\Cab{l}\in\ETLEMMAS{\vi}} \hspace{-.6cm}\Cabp{l}\
        \equiv\ \bot
    \end{eqnarray}
    Let $\vistarap{}\defas\exists \allB.\bigwedge_{\Cab{l}\in\ETLEMMAS{\vi}} \Cabp{l}$.
    Since the $\etaa{i}$s in \CTTA{\vi} are all total on \allalpha{}, then for each
    $\etaa{i}$, either $\etaap{i}\models \vistarap{}$ or $\etaap{i}\models
        \neg\vistarap{}$. The latter is not possible, because it would
    mean that  $\etaap{i}\wedge\vistarap{}\models\bot$, and hence, by
    \eqref{eq:teo3-hp3bool}, $\eta{}\in\ITTA{\vi}$, which would contradict the
    fact that $\CTTA{\vi}$ and $\ITTA{\vi}$ are disjoint.
    Thus we have:
    \begin{eqnarray}
        \label{eq:teo3-hp4}
        \bigvee_{\etaa{i}\in\CTTA{\vi}}\hspace{-.6cm}\etaap{i}\models\exists\allB.\bigwedge_{\Cab{l}\in\ETLEMMAS{\vi}}\hspace{-.6cm}
        \Cabp{l}.
    \end{eqnarray}



    \noi{} Hence, we have that:
    \begin{eqnarray}
        \label{eq:proof22}\nonumber
        &&
        \viap\ \wedge \exists\allB.\bigwedge_{\Cab{l}\in\ETLEMMAS{\vi}} \hspace{-.6cm}\Cabp{l}\
        \\
        \label{eq:proof32}\nonumber
        \mbox{{\em by \eqref{eq:decomposition}}:}\hspace{.2cm}    \hspace{-.3cm}&\equiv &\hspace{-.3cm}
        \left(\bigvee_{\etaa{i}\in\CTTA{\vi}}\hspace{-.6cm}\etaap{i}\ \vee\  \bigvee_{\rhoa{j}\in\ITTA{\vi}}\hspace{-.6cm}\rhoap{j}\right)
        \wedge \\
        &&\hspace{-.3cm}\exists\allB.\bigwedge_{\Cab{l}\in\ETLEMMAS{\vi}} \hspace{-.6cm}\Cabp{l}\
        \\
        \label{eq:proof42}\nonumber
        \mbox{{\em $\wedge/\vee$}:}\hspace{.2cm}   \hspace{-.3cm}&\equiv &\hspace{-.3cm}
        \left(\bigvee_{\etaa{i}\in\CTTA{\vi}}\hspace{-.6cm}\etaap{i}\ \wedge \exists\allB.\bigwedge_{\Cab{l}\in\ETLEMMAS{\vi}} \hspace{-.6cm}\Cabp{l}\right)
        \vee\\
        &&\hspace{-.3cm}\left(\bigvee_{\rhoa{j}\in\ITTA{\vi}}\hspace{-.6cm}\rhoap{j}
        \wedge \exists\allB.\bigwedge_{\Cab{l}\in\ETLEMMAS{\vi}} \hspace{-.6cm}\Cabp{l}\right)
        \\
        \label{eq:proof52}\nonumber
        \mbox{{\em by \eqref{eq:teo3-hp3bool}}:}\hspace{.2cm}   \hspace{-.3cm}&\equiv &\hspace{-.3cm}
        \left(\bigvee_{\etaa{i}\in\CTTA{\vi}}\hspace{-.6cm}\etaap{i}\ \wedge \exists\allB.\bigwedge_{\Cab{l}\in\ETLEMMAS{\vi}} \hspace{-.6cm}\Cabp{l}\right)
        \\
        \label{eq:proof62}\nonumber
        \mbox{{\em by \eqref{eq:teo3-hp4}}:}\hspace{.2cm}   \hspace{-.3cm}&\equiv &\hspace{-.3cm}
        \bigvee_{\etaa{i}\in\CTTA{\vi}}\hspace{-.6cm}\etaap{i}.
    \end{eqnarray}
\end{proof}

\paragraph{Proof of \Cref{teo:canonicity}}
\begin{proof}
    By applying \Cref{teo:teo-ext} to both $\via{}$ and
    $\viaprime{}$ we obtain:
    \begin{eqnarray}
        \label{eq:teo4-ext1}
        \viap{}  \wedge \exists\allB.\hspace{-.6cm}\bigwedge_{\Cab{l}\in\ETLEMMAS{\vi}}\hspace{-.6cm}\Cabp{l}&\equiv& \hspace{-.3cm}\bigvee_{\etaa{i}\in\CTTA{\vi}}\hspace{-.3cm}\etaap{i}
        \\
        \label{eq:teo4-ext2}
        \viaprimep{}  \wedge \exists\allBprime.\hspace{-.8cm}\bigwedge_{\Cabprime{l}\in\ETLEMMASPRIME{\viprime}}\hspace{-.8cm}\Cabprimep{l}&\equiv& \hspace{-.3cm}\bigvee_{\etaaprime{i}\in\CTTA{\viprime}}\hspace{-.3cm}\etaaprimep{i}.
    \end{eqnarray}
    By Property~\ref{prop:assignmentsets}(d), 
    $\via{}\Tequiv\viaprime{}$ if and only if
    $\CTTA{\vi} = \CTTA{\viprime}$, that is, if and only if:
    \begin{eqnarray}
        \label{eq:teo4-ext3}
        \bigvee_{\etaa{i}\in\CTTA{\vi}}\etaap{i}\equiv\bigvee_{\etaaprime{i}\in\CTTA{\viprime}}\etaaprimep{i}.
    \end{eqnarray}
    Thus, by combining \eqref{eq:teo4-ext3} with \eqref{eq:teo4-ext1} and \eqref{eq:teo4-ext2},
    we have that $\via{}\Tequiv\viaprime{}$ if and only if
    \eqref{eq:teo4-th-prop} holds.
\end{proof}

\paragraph{Proof of \Cref{teo:canonicity-technique}}
\begin{proof}
    Let \via{} and $\viaprime{}$ be \T-formulas.
    By \Cref{teo:canonicity}, $\via{}\Tequiv\viaprime{}$ if and
    only if~\eqref{eq:teo4-th-prop} holds.
    Since the \bdds{} are canonical and  \btot{} is injective,
    \eqref{eq:teo4-th-prop} holds if and only if
    $\Tbdd{}(\via{})=\Tbdd{}(\viaprime{})$.
\end{proof}

\section{Appendix: Extended Related Work}
\label{sec:apdx:rw}
\begin{table*}[t]
    \newcolumntype{L}[1]{>{\raggedright\let\newline\\\arraybackslash\hspace{0pt}}m{#1}}
    \newcolumntype{C}[1]{>{\centering\let\newline\\\arraybackslash\hspace{0pt}}m{#1}}
    \newcolumntype{R}[1]{>{\raggedleft\let\newline\\\arraybackslash\hspace{0pt}}m{#1}}
    \newcommand{\cmark}{\textcolor{green}{\ding{51}}}%
    \newcommand{\xmark}{\textcolor{red}{\ding{55}}}%
    \newcommand{\bmark}{*}%
    \newcommand{\rl}[2]{\parbox[t]{4mm}{\multirow{#1}{*}{\rotatebox[origin=c]{90}{#2}}}}%
    \centering
    \begin{tabular}{c|l|l|c|C{2cm}|C{1.5cm}|c}
                    & \textbf{Solver}                                                                                     & \textbf{Theory}    & \textbf{Avail.} & \textbf{Prune inconsistent paths} & \textbf{Semi-canonical} & \textbf{Canonical} \\
        \hline
        \rl{6}{BDD} & \textsc{EQ-BDDs}~\cite{frisogrooteEquationalBinaryDecision2000}                                     & \eq{}              & \bmark          & \cmark                            & \cmark                  & \xmark             \\
                    & \textsc{Goel-FM}~\cite{goelBDDBasedProcedures1998,goelBDDBasedProcedures2003}                       & \eq{}              & \xmark          & \cmark                            & \cmark                  & ?             \\
                    & \textsc{Goel-$e_{ij}$}~\cite{goelBDDBasedProcedures1998,goelBDDBasedProcedures2003}                 & \eq{}              & \xmark          & \xmark                            & \xmark                  & \xmark             \\
                    & \textsc{Bryant}~\cite{bryantBooleanSatisfiabilityTransitivity2002}                                  & \eq{}              & \xmark          & \cmark                            & \cmark                  & ?             \\
                    & \textsc{EUF-BDDs}~\cite{vandepolBDDRepresentationLogicEquality2005}                                 & \euf{}             & \bmark          & \cmark                            & \cmark                  & \xmark             \\
                    & \textsc{(0,S,=)-BDDs}~\cite{badbanAlgorithmVerifyFormulas2004,badbanZeroSuccessorEquality2005}      & $\eq\cup\set{0,S}$ & \xmark          & \cmark                            & \cmark                  & \xmark             \\
                    & \textsc{DDD}~\cite{mollerDifferenceDecisionDiagrams1999}                                           & \dl{}              & \xmark          & \cmark                            & \cmark                  & \xmark             \\
                    & \textsc{LDD}~\cite{chakiDecisionDiagramsLinear2009}                                                 & \laratint{}        & \cmark          & \xmark                            & \xmark                  & \xmark             \\
                    & \textsc{Chan}~\cite{chanCombiningConstraintSolving1997}                                             & \nla{}             & \xmark          & \cmark                            & \xmark                  & \xmark             \\
                    & \textsc{haRVey}~\cite{deharbeLightweightTheoremProving2003}                                         & any                & \xmark          & \cmark                            & \xmark                  & \xmark             \\
                    & \textsc{Fontaine}~\cite{fontaineUsingBDDsCombinations2002}                                          & any                & \xmark          & \cmark                            & \xmark                  & \xmark             \\
                    & \textsc{BDD+SMT}~\cite{cavadaComputingPredicateAbstractions2007,cimattiTighterIntegrationBDDs2010} & any                & *               & \cmark                            & \xmark                  & \xmark             \\
        \hline
        \rl{3}{SDD} &                                                                                                     &                    &                 &                                   &                                              \\
                    & \textsc{XSDD}~\cite{dosmartiresExactApproximateWeighted2019,kolbHowExploitStructure2020}           & \larat{}           & *               & \xmark                            & \xmark                  & \xmark             \\
                    &                                                                                                     &                    &                 &                                   &                                              \\
    \end{tabular}
    \caption{%
        Theory-aware decision diagram solvers. The column \emph{Available} indicates whether the solver is available for the given theory. The symbol ``\cmark{}'' means that the solver is publicly available, \bmark{} means that the solver is implemented within another tool and not directly usable, and \xmark{} means that the solver is not publicly available.
        In the column \emph{Canonical}, the symbol ``?'' indicates that the DDs may be canonical, but the authors do not provide a proof.
    }
    \label{tab:solvers}
\end{table*}

Several works have tried to leverage Decision Diagrams from the propositional to the \smt{} level.
Most of them are theory-specific, in particular focusing on \eq, \euf{} and (fragments of) arithmetic.
\euf{}-DDs are of particular interest in hardware verification~\cite{goelBDDBasedProcedures1998,goelBDDBasedProcedures2003}, while DDs for arithmetic have been mainly studied for the verification of infinite-state systems~\cite{chanCombiningConstraintSolving1997}.

In the following, we present an analysis of the most relevant works that leverage DDs from the propositional to the \smt{} level. We focus on generalization of \obdds{} and \sdds{}, as they are the most used DDs in the literature.
In Table~\ref{tab:solvers}, we summarize the main properties of the analyzed works.
From the table, we can see that most of the works are theory-specific, and while several \T-semicanonical DDs have been proposed, \T-canonical representations have been achieved only in some very-specific cases.
With the only exception of LDDs~\cite{chakiDecisionDiagramsLinear2009}, all the analyzed works do not have a public implementation or are implemented within other tools, making them not directly usable.

\paragraph{\Tbdds{} for \euf{}.}
In~\cite{goelBDDBasedProcedures1998,goelBDDBasedProcedures2003}, the authors describe two techniques to build \obdds{} for the theory of equality (\eq{}). The first 
consists in encoding each of the $n$ variables with $\lceil\log(n)\rceil$ bits, reducing to a Boolean formula. The resulting \obdd{} is, therefore, canonical, but its size is unmanageable even for small instances. The second approach consists in introducing a Boolean atom $e_{ij}$ for each equality $x_i=x_j$, and building a \obdd{} over these atoms. This essentially builds the \obdd{} of the Boolean abstraction of the formula, which allows for theory-inconsistent paths.
This problem has been addressed in~\cite{bryantBooleanSatisfiabilityTransitivity2002}, where transitivity lemmas are instantiated in advance and conjoined with the \obdd{}.
This approach is similar in flavour to our approach, and produces \obdds{} whose refinement is \euf-canonical; the main difference is that the procedure used to generate the lemmas is specific to the theory of equality, whereas our approach is general and can be applied to any theory supported by the SMT solver.

EQ-BDDs~\cite{frisogrooteEquationalBinaryDecision2000} extend \obdds{} to allow for nodes with atoms representing equation between variables.
EUF-BDDs~\cite{vandepolBDDRepresentationLogicEquality2005} extend EQ-BDDs to atoms involving also uninterpreted functions.
(0,S,=)-BDDs~\cite{badbanAlgorithmVerifyFormulas2004,badbanZeroSuccessorEquality2005} extend EQ-BDDs to atoms involving also the zero constant and the successor function.
In all three cases, rewriting rules are applied to prune inconsistent paths. The resulting \obdds{} are \T-semicanonical, but not \T-canonical.

%

\paragraph{\Tbdds{} for arithmetic.}

Difference Decision Diagrams (DDDs)~\cite{mollerDifferenceDecisionDiagrams1999} are a generalization of \obdds{} to the theory of difference logic (\dl{}). The building procedure consists in first building the refinement of the \obdd{} of the Boolean abstraction of the formula, and then pruning inconsistent paths by applying local and path reductions.
Local reductions are based on rewriting rules, leveraging implications between predicates to reduce redundant splitting. Path reductions prune inconsistent paths, both those going to the $\top{}$ and $\bot{}$ terminals.
The resulting DDD is \T-semicanonical, as \T-valid and \T-inconsistent formulas are represented by the $\top{}$ and $\bot{}$ DDDs, respectively. In general, however, they are not \T-canonical, even for formulas on the same atoms. Some desirable properties are discussed, and they conjecture that DDDs with these properties are canonical.  

LDDs~\cite{chakiDecisionDiagramsLinear2009} generalize DDDs to \laratint{} formulas. 
However, the implementation restricts to the theory of Two Variables Per Inequality (\tvpi{}) over real or integer variables. Moreover, only local reductions are applied, making them not even \T-semicanonical. Most importantly, not even contradictions are recognized.

In~\cite{chanCombiningConstraintSolving1997}, a procedure is described to build \Tobdds{} for nonlinear arithmetic (\nla). The procedure consists in building the refinement of the \obdd{} of the Boolean abstraction of the formula, and then using an (incomplete) quadratic constraint solver to prune inconsistent paths. As a result, the \Tobdd{} is not \T-semi-canonical, since \T-valid formulas may have different representations.

To the best of our knowledge, XSDDs~\cite{dosmartiresExactApproximateWeighted2019,kolbHowExploitStructure2020} are the only tentative to extend \sdds{} to support first-order theories. XSDDs have been proposed in the context of Weighted Model Integration (WMI), and extend \sdds{} by allowing for atoms representing linear inequalities on real variables in decision nodes. However, they only propose to refine the SDD of the Boolean abstraction of the formula, without any pruning of inconsistent paths. Simplifications are only done at later stages during the WMI computation.

\paragraph{\Tbdds{} for arbitrary theories.}


In~\cite{deharbeLightweightTheoremProving2003}, a general way has been proposed to build \Tobdds. The tool named \textsc{haRVey} first builds the refinement of the \obdd{} of the Boolean abstraction of the formula. Then, it looks for a \T-inconsistent path, from which it extracts a subset of \T-inconsistent constraints. The negation of this subset, which is a \T-lemma, is conjoined to the \Tobdd{} to prune this and possibly other \T-inconsistent paths. The procedure is iterated until no inconsistent paths are found.
Here, only the lemmas necessary to prune \T-inconsistent partial assignments satisfying the formula are generated, making the resulting \Tobdd{} not \T-semicanonical, as \T-valid formulas may have different representations.

The technique described in~\cite{fontaineUsingBDDsCombinations2002} is similar, but it generalizes to combination of theories.

In~\cite{cavadaComputingPredicateAbstractions2007}, the authors propose a general method to build \Tobdd{} by integrating an \obdd{} compiler with an \smtt{} solver, which is invoked to check the consistency of a path during its construction. The approach was refined in~\cite{cimattiTighterIntegrationBDDs2010}, where the authors propose many optimizations to get a tighter integration of the \smt{} solver within the \Tobdd{} construction. In both cases, all inconsistent paths are pruned, but the resulting \Tobdd{} is not \T-semicanonical.

%
%
%
\end{document}